\documentclass[journal]{IEEEtran}

\usepackage{cite}

\ifCLASSINFOpdf
   \usepackage[pdftex]{graphicx}
\else
\fi
\usepackage{amsmath}
\usepackage{dsfont}

\usepackage[utf8]{inputenc}
\usepackage[T1]{fontenc}
\usepackage{lmodern}
\usepackage{bm}
\usepackage{xcolor}

\usepackage{algorithm}
\usepackage{algpseudocode}

\begin{document}

\title{Data Transmission based on Exact Inverse Periodic Nonlinear Fourier Transform, Part I: Theory}

\author{Jan-Willem~Goossens,~\IEEEmembership{Student Member, IEEE,}
        Hartmut~Hafermann,~\IEEEmembership{Senior Member, IEEE}
        and~Yves~Jaou\"en~\IEEEmembership{}%
\thanks{J.-W.~Goossens and H.~Hafermann are with the Optical Communication Technology Lab, Paris Research Center, Huawei Technologies France, 92100 Boulogne-Billancourt, France}%
\thanks{J.-W.~Goossens and Y.~Jaou\"en are with LTCI, T\'el\'ecom Paris, Universit\'e Paris-Saclay, 91120 Palaiseau, France}%
}

\maketitle

\begin{abstract}
The nonlinear Fourier transform (NFT) decomposes waveforms propagating through optical fiber into nonlinear degrees of freedom, which are preserved during transmission. By encoding information on the nonlinear spectrum, a transmission scheme inherently compatible with the nonlinear fiber is obtained. Despite potential advantages, the periodic NFT (PNFT) has been studied less compared to its counterpart based on vanishing boundary conditions, due to the mathematical complexity of the inverse transform. In this paper we extract the theory of the algebro-geometric integration method underlying the inverse PNFT from the literature, and tailor it to the communication problem. We provide a complete algorithm to compute the inverse PNFT. As an application, we employ the algorithm to design a novel modulation scheme called nonlinear frequency amplitude modulation, where four different nonlinear frequencies are modulated independently. Finally we provide two further modulation schemes that may be considered in future research. The algorithm is further applied in Part II of this paper to the design of a PNFT-based communication experiment.
\end{abstract}

\section{Introduction}

\IEEEPARstart{N}{onlinear} Fourier transform (NFT) based communication is considered a promising route to address performance degradation of data transmission caused by nonlinear interference in optical fiber~\cite{Turitsyn2017}.
In contrast to mitigating or compensating nonlinear effects, they become an integral part of the signal design. Discovered in the sixties~\cite{Gardner1967}, the NFT --- also known as the inverse scattering transform in the mathematical physics context --- provides a tool to solve a wide class of nonlinear partial differential equations (the class of integrable equations~\cite{Das1989}) analytically. This includes the nonlinear Schr\"odinger equation (NLSE), which describes the space-time evolution of the envelope of the carrier in optical fiber~\cite{Agrawal2000}.
Even though signal evolution in the time-domain in presence of strong nonlinearity is complex, the evolution of the nonlinear spectrum of the NFT is simpler. When nonlinear degrees of freedom are modulated, their recovery at the receiver remains straightforward.

Most of the research so far has focused on the NFT in its conventional formulation with vanishing boundary conditions. In this approach, signals must decay sufficiently rapidly and be transmitted in burst mode, that is, be separated by zero-padding guard intervals to account for dispersion induced signal broadening.
Since its initial proposal~\cite{Yousefi2014,Pripelsky2014,Le2014}, the application of the NFT to the communication problem has seen progress on several fronts. This includes the development of techniques to precisely control the time-duration of a signal through b-modulation~\cite{Gui2018}, dispersion pre-compensation to reduce the size of the processing window at the receiver~\cite{Tavakkolnia2016} and the generalization to polarization division multiplexing~\cite{Goossens2017,Gaiarin2018,Civelli2018}. Feasibility of these ideas has also been demonstrated in numerous experiments~\cite{Le2017,Le2017b,Le2018,Aref2018,Gaiarin2018a,Gui2018,Gui2018a}, which improved aspects related to the digital signal processing, and culminated in the current record experimental data rate of 220 Gb/s~\cite{Yangzhang2019}.

A periodic version of the NFT (PNFT) has been developed from the mid seventies. The study of the periodic inverse scattering problem gives rise to the algebro-geometric approach to nonlinear integrable equations and is discussed in detail with example applications in physics in Ref.~\cite{Belokolos1994}. It reveals a close connection between Riemann surfaces and the theory of Abelian (or hyperelliptic) functions. These are generalizations of elliptic functions, and can be expressed as ratios of homogeneous polynomials\footnote{Polynomials whose nonzero terms all have identical degree.} of the Riemann theta function~\cite{Deconinck2004}. The connection between theta functions and nonlinear equations is discussed in Ref.~\cite{Dubrovin1981}.
The impact of the approach on modern mathematics and theoretical physics is described in Ref.~\cite{Matveev2008} with an emphasis on the historical context. Ref.~\cite{Osborne2010} gives an extensive account of nonlinear multi-dimensional Fourier analysis in the context of nonlinear ocean waves.

A first mention of the PNFT in the context of optical communication appears in Ref.~\cite{Yousefi2013}, where it is argued that based on the PNFT a discrete version of the NFT should be obtained. The PNFT has been suggested to be better suited for the communication problem by Wahls and Poor~\cite{Wahls2015}. Indeed, the PNFT can be considered the more natural generalization of the discrete Fourier transform --- which also assumes periodic boundary conditions --- to nonlinear channels. 
A communication scheme based on exact periodic solutions endowed with a cyclic prefix bears a close resemblance with orthogonal frequency-division multiplexing.

In the algebro-geometric approach~\cite{Belokolos1994,Kotlyarov2014,Tracy1988} to the inverse PNFT, exact solutions to the NLSE are given in terms of an analytical expression. This provides a number of potential advantages of the periodic over the conventional NFT.
The analytical form can, for example, be exploited to adjust certain properties of the solution, such as its temporal (or spatial) period, which we exploit in this paper. The underlying algebro-geometric structure may also provide a means to address the nonlinear multiplexing problem, that is, the computation of the superposition of two waveforms.
Algebro-geometric reduction~\cite{Belokolos1994}, where the \emph{symmetry} of the nonlinear spectrum allows one to reduce Abelian integrals and Riemann theta functions to lower genera, simplifies  computations~\cite{Smirnov2013}.
Further advantages that have been cited~\cite{Kamalian2016,Turitsyn2017} include a significant reduction of the peak to average power ratio compared to burst-transmission for PNFT symbols transmitted with a cyclic prefix. Furthermore, all information is encoded in a single period. Time-duration control is hence automatically built into the approach. While the cyclic prefix length depends on transmission distance, the processing window at the receiver is always equal to one period.

Stable fast algorithms to compute the nonlinear main and auxiliary spectra (forward PNFT) of a periodic waveform are available~\cite{Wahls2015,Kamalian2016} and a number of data transmission schemes based on modulating a small number of degrees of freedom have been proposed, such as transmission of perturbed plane-waves~\cite{Kamalian2016} or soliton-like pulses~\cite{Kamalian2018}. Algebro-geometric reduction has been exploited in the communication context in Ref.~\cite{Kamalian2018c} for genus 2. The generalization to higher genus however is not straightforward. The first experimental demonstration of PNFT-based data transmission was presented only recently~\cite{Goossens2019}. 
A complementary method for the numerical computation of periodic solutions (inverse PNFT) is based on the solution of an associated Riemann-Hilbert problem~\cite{Kamalian2018} to obtain periodic solutions numerically. The advantage is a reduction of the computational complexity compared to the algebro-geometric approach employed in this paper, while the algebro-geometric approach can leverage advantages associated with the availability of a solution in analytical form.

The numerical computation of solutions within the algebro-geometric approach has been partially  addressed in the literature. An algorithmic approach to the computation of period matrices of Riemann surfaces was presented in Ref.~\cite{Deconinck2001}. An overview of further computational approaches to Riemann surfaces can be found in Ref.~\cite{Bobenko2011}. In~\cite{Kalla2012} a numerical approach to the computation of solutions for real algebraic curves (which correspond to the spectra of the defocusing NLSE) has been presented.
Numerical aspects of the evaluation of Riemann theta functions has previously been studied~\cite{Deconinck2004} and efficient software packages are available~\cite{Swierczewski2016}.

In this paper, we develop the algebro-geometric approach to the inverse PNFT for the application to the optical communication problem. 
At the core of the paper, we develop a fully automatized procedure to compute the exact solutions based on the algebro-geometric integration method. 
To the best of our knowledge, this is the first complete algorithm for the inverse PNFT which, starting from a main spectrum, yields all parameters required for the evaluation of the analytical solution in terms of integrals over a Riemann surface.
We present an algorithm to obtain a complete set of closed integration paths (the basis of the homology group, or homology basis for short) on the two-sheeted Riemann surface, and a formula to evaluate integrals over these paths. Benchmark results for the inverse PNFT are also provided.
As applications, we discuss three different modulation schemes based on the approach.
In Part II of this paper~\cite{GoossensII}, the obtained algorithm for the inverse PNFT is utilized to design a constellation for the experimental demonstration of PNFT-based transmission.

Transmission schemes based on the PNFT so far are less developed than for the conventional counterpart. To large extent this is due to the mathematical complexity of the inverse transform. To reduce this hurdle, we have extracted the theory from the specialized mathematical physics literature and formulated it in a unified notation suitable to the communication problem. 
In the Appendix we briefly introduce the mathematical notions underlying the algebro-geometric approach and provide a detailed outline of the derivation with links to the relevant references. 
Due to space constraints, the derivation can neither be a complete nor a mathematically rigorous exposition including all necessary proofs, which would also be largely redundant. Instead the purpose is to sketch the derivation of the formulas required for our work and to provide the reader with a means to navigate the specialized literature.

The paper is organized as follows: In Section~\ref{sec:pnft} we introduce the mathematical structure behind the PNFT and the definition of the forward transform. The inverse PNFT is introduced in Section~\ref{sec:ipnft}. Section~\ref{sec:algorithm}
details the algorithm for the automatic computation of the inverse PNFT. Section~\ref{sec:applications} provides three different applications of the algorithm before we conclude the paper in Section~\ref{sec:conclusion}.

\subsection*{Notation}

In this paper we use boldface $\bf{A}$ to denote $g\times g$-matrices while $A_{i,j}$ denotes the element of  $\bf{A}$ in 	row $i$ and column $j$. We use underlining, $\underline{V}$, to denote a $g$-dimensional column vector, where $V_j$ is the $j$-th component of that vector. Differentials are always recognizable by the Leibniz-notation, $d\psi$. The complex conjugate of a complex number $z$ is denoted $\bar{z}$, while its real and imaginary part are denoted by $\Re(z)$ and $\Im(z)$, respectively.

\section{Periodic NFT}
\label{sec:pnft}

We formulate the PNFT for the focusing nonlinear Schr\"odinger equation (NLSE)~\cite{Turitsyn2017}, which describes the evolution of light in anomalous dispersion optical fiber and is given in dimensionless form by
\begin{equation}
i\frac{\partial q(t,z)}{\partial z} + \frac{\partial^2 q(t,z)}{\partial t^2} + 2|q(t,z)|^2 q(t,z) = 0.
\label{eq:nlse}
\end{equation}
Contrary to most of literature, where the roles of space and time are reversed, the above notation is suitable to the optical channel.
The NLSE is an integrable nonlinear differential equation.
Integrability hinges on the fact that the NLSE arises as the compatibility condition $\partial^2 \Phi(t,z,\lambda)/(\partial t \partial z) = \partial^2 \Phi(t,z,\lambda)/(\partial z \partial t)$ between two \emph{linear} partial differential equations~\cite{Lax1968}:
\begin{align}
\frac{\partial \Phi}{\partial t} &=  \left[-i\lambda\sigma_3 + \begin{pmatrix}0& q(t,z)\\-\bar{q}(t,z)&0\end{pmatrix}\right]\Phi=\!\!\mathop:U(t,z,\lambda) \Phi,\nonumber\\
\frac{\partial \Phi}{\partial z} &= \left[-2\lambda U + \begin{pmatrix} -iq(t,z)\bar{q}(t,z)&-\frac{\partial q(t,z)}{\partial t}\\-\frac{\partial \bar{q}(t,z)}{\partial t}&iq(t,z)\bar{q}(t,z),\end{pmatrix}\right]\Phi,\nonumber\\
\sigma_3 &= \begin{pmatrix}1&0\\0&-1\end{pmatrix}, \Phi(0,0,\lambda) = \begin{pmatrix}1&0\\0&1\end{pmatrix}.\label{eq:UV}
\end{align}
These equations can be understood as a scattering problem in which $q(t,z)$ takes on the role of a potential. In the context of the algebro-geometric approach, the literature often formulates the PNFT for the so-called coupled nonlinear Schr\"odinger equation~\cite{Tracy1984, Belokolos1994} (not to be confused with the Manakov equation), of which the focusing NLSE is a special case.

For a time-periodic signal with period $T$, $q(t,z)=q(t+T,z)$, the columns of $\Phi$ become Bloch functions\footnote{In solid-state physics, Bloch functions arise as eigenfunctions in the treatment of periodic potentials~\cite{Kittel1996}.}, and the main spectrum is given by those values $\lambda_j$ for which the Bloch functions become (anti-)periodic. A different way to state this is that the main spectrum of the PNFT is given by those $\lambda_j$ for which the monodromy matrix $M$,
\begin{equation}
M(t,z,\lambda) := \Phi(t+T,z,\lambda)\Phi^{-1}(t,z,\lambda),\label{eq:monodromy}\end{equation} has eigenvalues $\pm 1$~\cite[App. A]{Tracy1988}. The monodromy matrix $M$ relates a solution of Eq.~\eqref{eq:UV} at time $t$ to a solution after precisely one period at time $t+T$. 

The main spectrum always consists of complex conjugate pairs because the operator $U(t,z,\lambda)$ is skew-Hermitian~\cite{Kotlyarov2014}. It can therefore also be described by the $g+1$ points in the spectrum with positive imaginary part. When $\Im(\lambda_j)=0$, the eigenvalue is a double point, which does not contribute to the dynamics of $q(t,z)$~\cite{Wahls2015}. The dynamics of the wave are captured by the time- and space-dependent auxiliary spectrum. It is determined by the positions $\mu_j(t,z)$  where the off-diagonal element $M_{1,2}$ of the monodromy-matrix vanishes: $M_{1,2}(t,z,\mu_j(t,z))=M_{2,1}(t,z,\bar{\mu}_j(t,z)) = 0$~\cite[above (5.2)]{Kotlyarov2014}.

This definition is due to Kotlyarov and Its~\cite{Kotlyarov2014}. The definition of the auxiliary variables is not unique. An alternative set of auxiliary variables was given by Ma and Ablowitz~\cite{Ma1981}. This description simplifies the retrieval of $q(t,z)$ from the auxiliary parameters, at the cost of a more involved evolution of the auxiliary spectrum itself. However, no condition seems to be known to ensure that such a set of auxiliary parameters corresponds to a solution of the NLSE~\cite{Wahls2015}. We will therefore not consider this alternative set of auxiliary variables.

\subsection*{Finite-gap solutions}

When the main spectrum of a waveform consists of only a finite number of non-degenerate points, the waveform is referred to as a finite-gap or finite-band solution. For finite-gap solutions, $q(t,z)$ and the auxiliary spectrum $\mu_j(t,z)$ are described by a set of coupled partial differential equations:
\begin{align}
\frac{\partial \mu_j(t,z)}{\partial t} &= \frac{-2i\sigma_j\sqrt{\prod_{k=1}^{2g+2}(\lambda_k-\mu_j(t,z))}}{\prod_{l\neq j}(\mu_j(t,z)-\mu_l(t,z))},\label{eq:derivative_of_mu}
\end{align}
\begin{align}
\frac{\partial \log q(t,z)}{\partial t} =2i\left(\sum_{j=1}^{g}\mu_j(t,z)-\frac{1}{2}\sum_{k=1}^{2g+2}\lambda_k\right),\label{eq:derivative_of_q}
\end{align}
 with similar, somewhat more complicated equations for the $z$-derivatives (see Ref.~\cite{Kotlyarov2014} and Eqs.~\eqref{eq:app:UV} and~\eqref{eq:app:derivative_of_q_app} in  the Appendix). The differential equations for $\mu_j(t,z)$ are derived from the evolution of the monodromy matrix, Eq.~\eqref{eq:monodromy}, which in turn is determined by the scattering problem~\eqref{eq:UV}. Although the differential equations are derived for periodic solutions, it turns out that they also have solutions
that are not periodic. It has been shown that these differential equations can be taken as the defining equations of finite-gap solutions~\cite{Tracy1984}, which are not necessarily periodic.

For given main spectrum $\lambda_j$, the solution to these equations is fully specified by the initial condition $\mu_j(0,0)$ and $q(0,0)$. The absolute value of $q(0,0)$ is determined by the initial condition for the auxiliary spectrum, while the phase is arbitrary. Not every choice of auxiliary spectrum corresponds to a solution of the NLSE. A sufficient constraint to guarantee an initial condition for the auxiliary spectrum $\mu_j(0,0)$ to yield a solution is given in Appendix~\ref{app:aux_spec}. It was first realized by Kotlyarov and Its~\cite{Kotlyarov2014} that the auxiliary variables $(\mu_j(t,z),\sigma_j(t,z))$ should be interpreted as points on a Riemann surface~\eqref{eq:riemann_surface}.

\section{Exact inverse PNFT}
\label{sec:ipnft}
This section introduces the algebro-geometric approach to the inverse PNFT. The presentation requires certain mathematical notions which are detailed in Appendix~\ref{app:math}. Readers not familiar with these notions are recommended to read Appendix~\ref{app:math} first.

The inverse PNFT is a mapping from the main spectrum $\lambda_k$ and initial conditions for the auxiliary spectrum $\mu_j(0,0)$  to a solution of the NLSE $q(t,z)$.
As outlined in the Appendix, the integration of the partial differential equations for $\mu_j(t,z)$ and $q(t,z)$ leads to exact solutions of the following form:
\begin{equation}
q(t,z) = K_0\frac{\theta\left(\frac{1}{2\pi}(\underline{\omega}t+\underline{k}z+\underline{\delta}^-)|\bm{\tau}\right)}{\theta\left(\frac{1}{2\pi}(\underline{\omega}t+\underline{k}z+\underline{\delta}^+)|\bm{\tau}\right)} e^{i\omega_0 t+ik_0z},\label{eq:thetasol}
\end{equation}
where the Riemann theta function is defined by:
\begin{equation}
\theta(\underline{x}|\bm{\tau}) = \sum_{m_1=-\infty}^\infty \ldots\sum_{m_g=-\infty}^\infty \exp(\pi i\underline{m}^T\bm{\tau}\underline{m}+ 2\pi i\underline{m}^T\underline{x}).
\label{eq:theta}
\end{equation}
It is periodic in all components of the vector $\underline{x}$ with period 1. 
$\bm{\tau}$ is the period matrix. It is symmetric and has positive definite imaginary part, which guarantees convergence of the series in the theta function.
All parameters on the right-hand side of Eq.~\eqref{eq:thetasol} are obtained as integrals over the Riemann surface $\Gamma$ defined by
\begin{equation}
\Gamma: \left\{(P,\lambda)\textrm{, }P^2 = \prod_{k=1}^{g+1} (\lambda-\lambda_k)(\lambda-\bar{\lambda}_k), P,\lambda\in\mathds{C}\right\}.
\label{eq:riemann_surface}
\end{equation}
In particular, the parameters $\omega_j$ and $k_j$ are given by
\begin{align}	
\omega_j &= -4\pi i (\bm{A}^{-1})_{j,g},\label{eq:periods}\\ 
k_j &= -8\pi i \left[(\bm{A}^{-1})_{j,g-1}+\frac{1}{2}(\bm{A}^{-1})_{j,g}\left(\sum_{k=1}^{2g+2}\lambda_k\right)\right],\label{eq:periods2}
\end{align}	
where 
$(\bm{A}^{-1})_{j,g-1}:= 0$ for $g<2$ and $\bm{A}$ is a $g\times g$ matrix determined by
\begin{equation}
A_{jk} = \int_{a_k} dU_j,\quad j,k =1,\ldots,g.\label{eq:Amat}
 \end{equation}
The $dU_j$ are the basis of holomorphic differentials defined in~\eqref{eq:holomorphic_differentials} in Appendix~\ref{app:math}.
The integration paths $a_j$ form half of the canonical homology basis on the Riemann surface.

In terms of the periods of the holomorphic differentials over the surface $\Gamma$ the period matrix is defined as:
\begin{equation}
\bm{\tau} = \bm{A}^{-1}\bm{B}\textrm{,\,\,\, } B_{jk} = \int_{b_k} dU_j,
\label{eq:tau}
\end{equation}
where the cycles $b_j$ form the other half of the homology basis.
The parameters $\underline{\delta}^-$ and $\underline{\delta}^+$ are the only ones dependent on the auxiliary spectrum, and are defined as:
\begin{align}
\frac{1}{2\pi}\delta^\pm_j =& \int_{p_0}^{\infty^\pm}d\psi_j -\frac{1}{2}\tau_{jj} + \sum_{k=1}^g\int_{a_k}d\psi_k(p')\int_{p_0}^{p'}d\psi_j(p)
\nonumber\\&- \sum_{k=1}^g\int_{p_0}^{\mu_k(0,0)}d\psi_j\label{eq:deltapm}.
\end{align}
Here $p_0$ is an arbitrary base point on the Riemann surface. 
To ensure the integrals in this definition are uniquely defined, none of the paths for the integrals must cross any of the $a$- or $b$-cycles.
Here crossing means that the path cannot be deformed continuously to avoid the crossing without crossing a branch point.
The holomorphic differentials $d\underline{\psi}$ are defined by
\begin{equation}
d\underline{\psi} = \bm{A}^{-1}d\underline{U}.\label{eq:normdiff}\end{equation}
They form a homology basis which is normalized: their periods over the $a$-cycles give the identity matrix: $\delta_{j,k} = \int_{a_k} d\psi_j$. 
In terms of the normalized differentials, the period matrix is given by the $b$-periods: $\tau_{j,k} = \int_{b_k} d\psi_j$.
It was proven~\cite[Eq. 4.3.22]{Belokolos1994} that each valid initial condition for the auxiliary spectrum corresponds to a vector $\underline{\delta}^+$ with $\Im(\underline{\delta}^+)=0$ and vice versa. Therefore choosing a vector $\underline{\delta}^+$ implicitly fixes the initial condition for the auxiliary spectrum (see also Appendix~\ref{app:integrateq}). This condition is simpler to implement than the constraint on the auxiliary spectrum detailed in Appendix~\ref{app:aux_spec}. We exploit this in the phase modulation scheme introduced below.

Note that $\delta^-_j$ and $\delta^-_j + 2\pi$ describe the exact same solution due to the periodicity of the theta function.
We choose $\delta_j^+=0$ and obtain $\delta_j^-$ from:
\begin{equation}
\frac{1}{2\pi}(\delta^+_j -\delta^-_j) =\int_{\infty^-}^{\infty^+}d\psi_j,
\label{eq:delta}
\end{equation} 
where the path from $\infty^-$ to $\infty^+$ must not cross any of the cycles in the homology basis. 

To solve the differential equations for $\mu_j(t,z)$ and $q(t,z)$  we must also provide an initial value $q(0,0)$. The amplitude $|q(0,0)|$ is fixed by the implicit choice of auxiliary spectrum, while the phase of $q(0,0)$, and that of $q(t,z)$ in general, is a free parameter~\cite{Kotlyarov2014}.
This can be seen from the fact that Eq.~\eqref{eq:derivative_of_q} only provides the rate of change of $q$. When $q(t,z)$ is a solution to the NLSE~\eqref{eq:nlse}, $q(t,z)e^{i\phi}$ is also a solution. Similarly, the phase of $K_0$ in Eq.~\eqref{eq:thetasol} is a free parameter.

Finally, the parameters $|K_0|$, $\omega_0$ and $k_0$ are obtained as the subleading terms in an expansion of meromorphic differentials $d\Omega_j$, $j = 0,1,2$ in $\lambda$ around the point $\lambda=\infty$~\cite[Eq. 4.3.6]{Belokolos1994}. They are uniquely defined by the following properties (see Appendix~\ref{app:baker} and~\cite[p. 91]{Belokolos1994}):
\begin{enumerate}
\item The differentials are normalized: $\int_{a_j} d\Omega_k = 0$.
\item Their asymptotic behavior is given by: \begin{align}
\!\!\!\!\!\!\!\Omega_0(p) &= \pm(\log \lambda-\frac{\log(-\frac{1}{4}|K_0|^2)}{2} + \mathcal{O}(\lambda^{-1})),\, p\to\infty^\pm,
\nonumber\\
\!\!\!\!\!\!\!\Omega_1(p) &= \pm(\lambda + \frac{\omega_0}{2}+\mathcal{O}(\lambda^{-1})),\, p\to\infty^\pm,
\nonumber\\ 
\!\!\!\!\!\!\!\Omega_2(p) &= \pm(2\lambda^2+\frac{k_0}{2} + \mathcal{O}(\lambda^{-1})),\, p\to\infty^\pm.
\label{eq:OmegaAsymptotics}
\end{align}
\item $d\Omega_k$  has no other singularities.	
\end{enumerate}The integral of $d\Omega_k$ from $\infty^-$ to $\infty^+$ is given by the difference between the limits: 
\begin{equation}
\int_{\infty^-}^{\infty^+}d\Omega_k = \lim_{p\to\infty^+}\Omega_k(p)-\lim_{p\to\infty^-}\Omega_k(p).
\end{equation}
Therefore, the subleading terms are formally determined by (see Appendix~\ref{app:baker}):
\begin{align}
-\log(-\frac{1}{4}|K_0|^2)&= \int^{\infty^+}_{\infty^-} d\Omega_0 - 2\int_1^\infty \frac{1}{\lambda} d\lambda,\label{eq:evalomega0}\\
\omega_0 &= \int^{\infty^+}_{\infty^-} d\Omega_1-2\int_0^\infty d\lambda,\label{eq:evalomega1}\\
k_0 &=\int^{\infty^+}_{\infty^-} d\Omega_2-2\int_0^\infty 4\lambda d\lambda.\label{eq:evalomega2}
\end{align}
The last integral in each equation subtracts the leading divergence from the first. Its lower boundary is determined such that it does not introduce an additional constant term.

\begin{figure*}[t]
\begin{center}
 \includegraphics[width=0.3\textwidth]{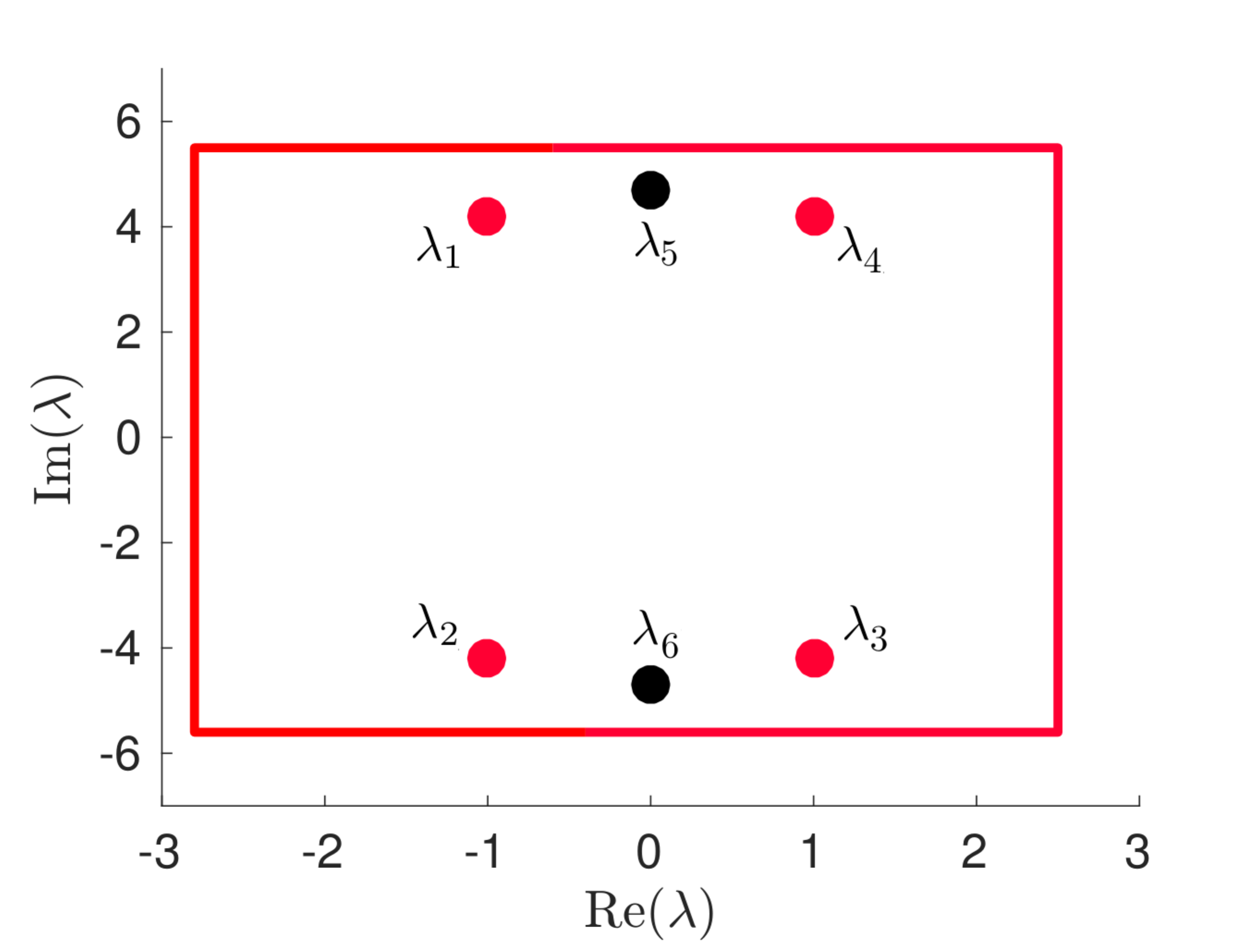}\includegraphics[width=0.3\textwidth]{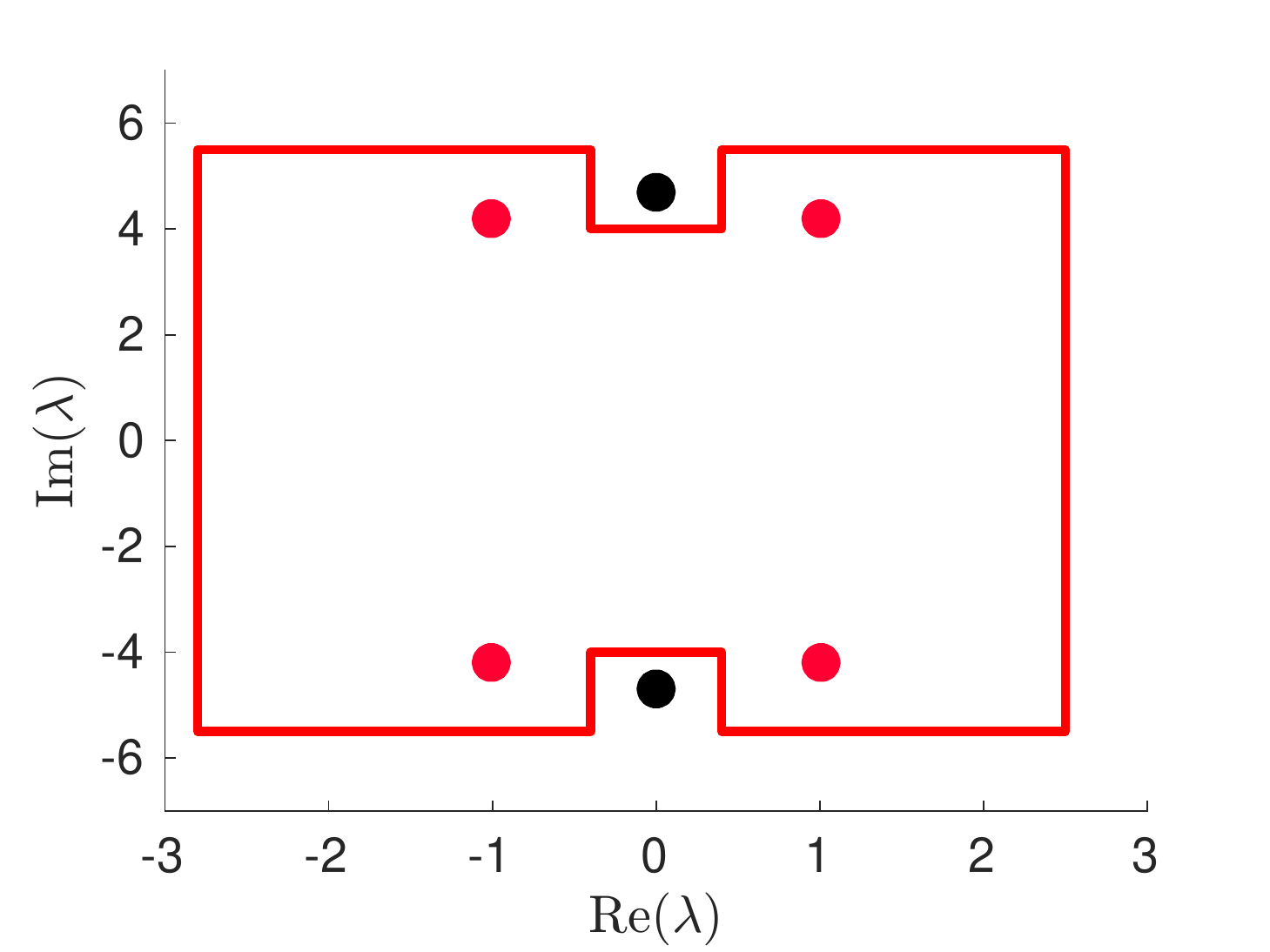}\includegraphics[width=0.3\textwidth]{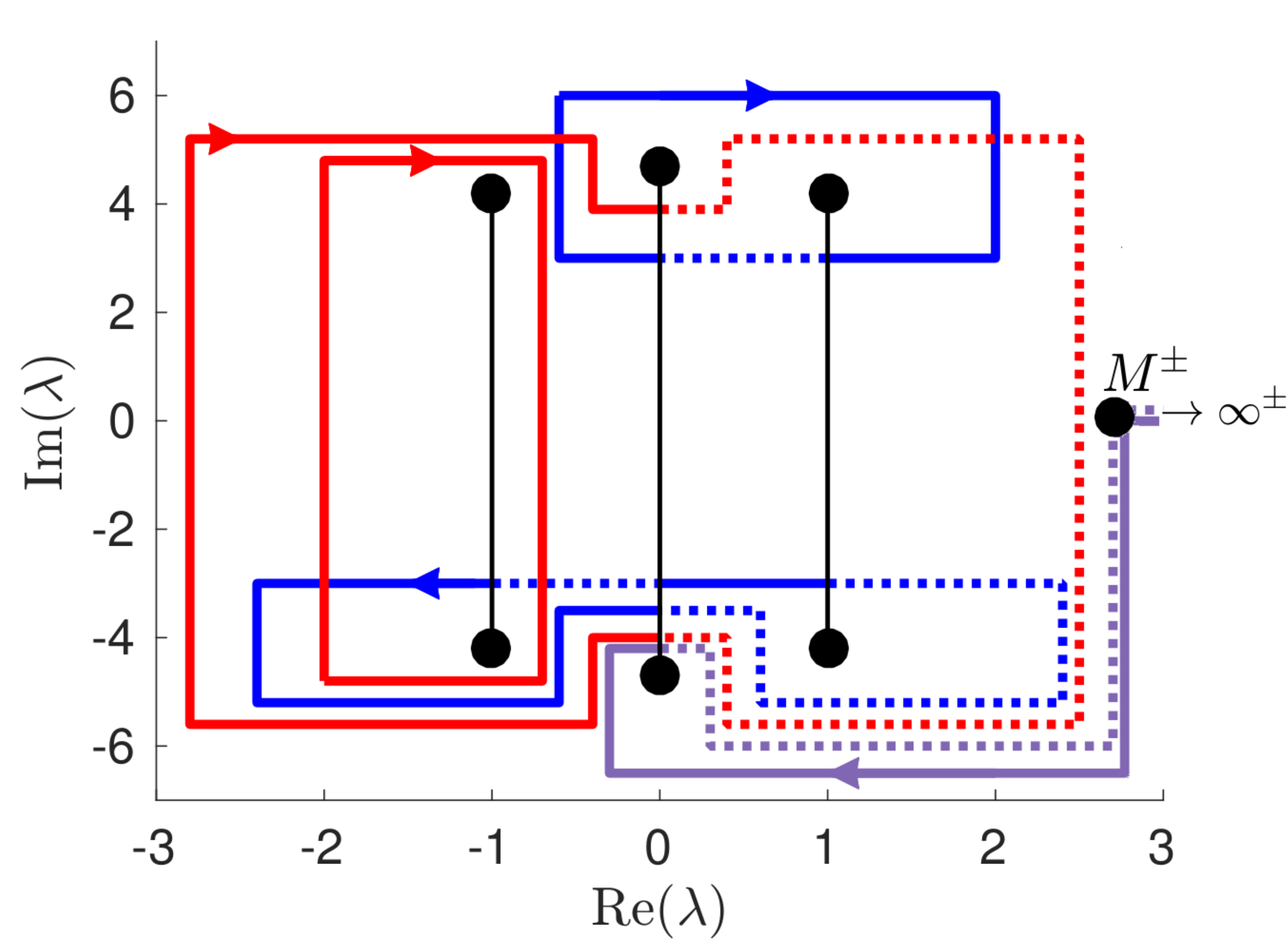}
\end{center}
 \caption{\label{fig:cycle_algo} Three different steps in the construction of a homology basis. Left: The initial rectangle constructed for a path around the red eigenvalues. The black eigenvalues ended up on the inside of the cycle accidentally. Middle: The same path, after the cut-out procedure. Right: The full homology basis, including the sheet changes and directions of the different cycles. The dashed and solid parts of the same cycle lie on the different sheets of the Riemann surface. The path to and from $\infty^\pm$ is also shown. The location of the branch cuts (vertical black lines) is illustrative and depends on the definition of the square root.}
\end{figure*}

\section{Inverse PNFT algorithm}
\label{sec:algorithm}
In this section we provide an algorithm to numerically compute a solution in the form of Eq.~\eqref{eq:thetasol}. The input of this algorithm are the main spectrum\footnote{Note that we technically only need to supply half of the spectrum, since the spectrum comes in complex conjugate pairs.}, $\{\lambda_{j}|j=1,\ldots, 2g+2\}$, and a real phase vector $\underline{\delta}^+$. The output is the solution $q(t,z)$ corresponding to that main spectrum and choice of $\underline{\delta}^+$. The computation of the PNFT is divided into the following subtasks:
First, obtain a homology basis for the Riemann surface, as well as a path to $\infty^\pm$. Next, compute the parameters defined by~\eqref{eq:periods},~\eqref{eq:periods2},~\eqref{eq:tau} and~\eqref{eq:delta} by numerically integrating their defining integrals over these paths. 
Then evaluate the integrals over the meromorphic differentials and isolate their subleading behavior to obtain $|K_0|$, $\omega_0$ and $k_0$ according to Eqs.~\eqref{eq:evalomega0} to \eqref{eq:evalomega2}.
Finally, given all parameters, $q(t,z)$ must be evaluated. The evaluation of the Riemann theta function has been studied in literature~\cite{Deconinck2004} and efficient software packages are available~\cite{Swierczewski2016}. We therefore only provide algorithms for the other three steps.

\subsection{Computing integration paths}
\label{sec:homology_basis}

The first step in obtaining the theta function parameters is the computation of a canonical homology basis. A homology group element (cycle) of the Riemann surface is completely determined by stating which branch points $\lambda_j$ are inside it.
By taking the following sets as points inside the cycle for the different basis elements, a homology basis is obtained:
\begin{equation}
a_j: \{\lambda_1,\cdots,\lambda_{2j}\}, b_j: \{\lambda_{2j}, \lambda_{2j+1}\}, \quad j=1,\ldots,g.\label{eq:cycles}
\end{equation}
To simplify the construction of these cycles, we organize the $\lambda_j$ in an order such that
\begin{align}\lambda_{2j-1}=\bar{\lambda}_{2j},\ j&=1,\ldots,g+1,\label{eq:sort1}
\\|\Im(\lambda_{2j})|\leq|\Im(\lambda_{2j+1})|,\ j&=1,\ldots,g,\label{eq:sort2}\\
\textrm{sgn}(\Im(\lambda_{2j})) = \textrm{sgn}\Im((\lambda_{2j+1})),\ j&=1,\ldots,g\label{eq:sort3}.
\end{align}
The ordering is illustrated for $g=2$ in the left panel of Fig.~\ref{fig:cycle_algo}.
We construct the individual elements in the homology basis by a 'cut-out' procedure. We first determine a cycle in the form of a rectangle that contains all necessary points, and then cut out any points from the spectrum that were supposed to be outside the cycle, but which ended up on the inside of the original rectangle. We represent a path $\gamma$ as an ordered sequence of $N$ waypoints $[\nu_1,\ldots, \nu_N]$ of the $\lambda$-coordinate in the complex plane. 
The complete description requires the location of the sheet-changes, as described in Sec.~\ref{sec:sheetchanges} below.
The path is obtained by connecting subsequent points by straight lines. A closed path has $\nu_1=\nu_N$.

As an example we construct a cycle around the points in the set $\tilde{\lambda}^{\text{in}}$ while ensuring that all other branch points are excluded. To maximize the accuracy of the numerical integration over the homology basis, we attempt to stay as far as possible from the branch points, since at these points the integrands have a singularity. We define $\epsilon =\frac{1}{2} \min_{i,j} \left(\max (|\Re(\lambda_i-\lambda_j)|,|\Im(\lambda_i-\lambda_j|))\right)$, which is chosen such that a square with side length $2\epsilon$ centered around one branch point has a distance of at least $\epsilon$ to any other.

First, we find the higher ($H$) and lower ($L$) extreme values in real ($R$) and imaginary ($I$) part:
\begin{align}
L_R = \min_j \Re(\lambda_j)-\epsilon,\quad H_R &= \max_j \Re(\lambda_j)+\epsilon,\label{eq:Re}\\
L_I = \min_j \Im(\lambda_j^{\text{in}})-\epsilon,\quad
H_I &= \max_j \Im(\lambda_j^{\text{in}})+\epsilon.\label{eq:Im}
\end{align}
These values define the corners of a rectangle: $[L_R+iH_I, H_R+iH_I, H_R+iL_I, L_R+iL_I]$, which is guaranteed to contain all points $\tilde{\lambda}^{\text{in}}$. Note that the left and right boundaries are chosen such that they enclose \emph{all} points. The rectangle can also contain additional points.
By virtue of~\eqref{eq:cycles} and~\eqref{eq:sort2}, the cycles $a_j$ contain the $2j$ smallest eigenvalues by imaginary part. The points to be excluded are therefore already outside the cycle or guaranteed to lie inside a band of width $\epsilon$ at the top or bottom of the rectangle. By virtue of~\eqref{eq:cycles} and~\eqref{eq:sort3}, a $b$-cycle is always located on one side of the real axis. Because of~\eqref{eq:sort2}, we have for the imaginary part of all other branch points $\lambda_i$ in $b_j$ that $\Im\lambda_i \leq \Im\lambda_{2j}$ or $\Im\lambda_i \geq \Im\lambda_{2j+1}$. The same conclusion therefore holds for the $b$-cycles.
Assuming the cycle is traversed clockwise starting from the top left corner, a point $\lambda \notin \tilde{\lambda}^\text{in}$ near the top is excluded from the cycle by adding the extra waypoints
\begin{align}
[&(\Re(\lambda)-\epsilon)+iH_I,\, (\Re(\lambda)-\epsilon)+i(\Im(\lambda)-\epsilon),\nonumber\\ 
 &(\Re(\lambda)+\epsilon) + i(\Im(\lambda)-\epsilon),\, (\Re(\lambda)+\epsilon) + iH_I)],\nonumber
\end{align}
which draw a small rectangular path around the point $\lambda$ at a distance $\epsilon$.
An example is provided in the first two images in Fig.~\ref{fig:cycle_algo}. If a point outside the cycle lies closer than $\epsilon$ to the boundary, a similar cut-out procedure can be applied to increase the distance of the path to the point. This can improve the accuracy of the numerical evaluation of the differentials, since they diverge at the branch points $\lambda_j$. 

The pseudocode for the algorithm is shown in Algorithm~\ref{alg:homology}. The tilde denotes arrays. The input is an array of points ordered according to~\eqref{eq:sort1} --~\eqref{eq:sort3}.
The output are arrays for for the $a$- and $b$-cycles in the homology basis, each containing an ordered set of waypoints.
The procedure \verb|GetCycle| in Algorithm~\ref{alg:cycle} returns a cycle as an array of waypoints given the array of branch points to be inside the cycles, the array of all branch points and the constant $\epsilon$.
It first determines the points of the enclosing rectangle and arrays of points to be excluded at the top and bottom. These points are cut out one by one in order.
Note that according to lines 10 and 13 eigenvalues that are outside the rectangle but within a distance $\epsilon$ from its borders, are also excluded.
To simplify the determination of the direction of the cycles later, it proves useful that they only meet in points, and never over extended segments.
An overlap of segments between two cycles can be avoided by scaling $\epsilon$ by a factor slightly smaller than one and unique to each cycle when drawing it (lines 2--5, 23--25 and 30--32 in Algorithm~\ref{alg:cycle}), while keeping the original value of $\epsilon$ when deciding which points to exclude (lines 10, 13).

\begin{algorithm}[t]
\caption{Construction of homology basis}\label{alg:homology}
\begin{algorithmic}[1]
\Procedure{HomologyBasis}{$\tilde{\lambda}$}
\State $\epsilon \gets \frac{1}{2}\min_{k,l}\max (\Re(\tilde{\lambda}[k]-\tilde{\lambda}[l]), \Im(\tilde{\lambda}[k]-\tilde{\lambda}[l]))$
\State $\tilde{a} \gets \{\}$
\State $\tilde{b} \gets \{\}$
\For{$j = 1,\, \ldots, g$}	 :
\State $\lambda^{\text{in}}_A \gets \{\tilde{\lambda}[1],\ldots, \tilde{\lambda}[2j]\}$\Comment{Eq.~\eqref{eq:cycles}}
\State $\lambda^{\text{in}}_B \gets \{\tilde{\lambda}[2j], \tilde{\lambda}[2j+1]\}$\Comment{Eq.~\eqref{eq:cycles}}
\State append($\tilde{a}, \text{GetCycle}(\tilde{\lambda}^{\text{in}}_A, \tilde{\lambda}, \epsilon)$)
\State append($\tilde{b}, \text{GetCycle}(\tilde{\lambda}^{\text{in}}_B, \tilde{\lambda}, \epsilon)$)
\EndFor
\State\Return{$\{\tilde{a}, \tilde{b}\}$} \Comment{Return lists of waypoints}
\EndProcedure
\end{algorithmic}
\end{algorithm}

\begin{algorithm}
\caption{Construction of closed cycle}\label{alg:cycle}
\begin{algorithmic}[1]
\Procedure{GetCycle}{$\tilde{\lambda}^\text{in}$, $\tilde{\lambda}$, $\epsilon$}
\State $H_R \gets \max_k(\Re(\tilde{\lambda}[k]))+\epsilon$,   \Comment{Eq.~\eqref{eq:Re}}
\State $L_R \gets \min_k(\Re(\tilde{\lambda}[k]))-\epsilon$ \Comment{Eq.~\eqref{eq:Re}}
\State $H_I \gets \max_k(\Im(\tilde{\lambda}^\text{in}[k]))+\epsilon$ \Comment{Eq.~\eqref{eq:Im}}
\State $L_I \gets \min_k(\Im(\tilde{\lambda}^\text{in}[k]))-\epsilon$ \Comment{Eq.~\eqref{eq:Im}}
\State$\tilde{\lambda}^{\text{xtop}} \gets \{\}$
\State$\tilde{\lambda}^{\text{xbot}} \gets \{\}$
\For{$k = 1,\, \ldots, 2g+2$}	
\If{$\tilde{\lambda}[k]$ $\notin$ $\tilde{\lambda}^\text{in}$}
\If{\!$\Im(\tilde{\lambda}[k])\!\geq\! H_I\! -\! \epsilon$\! and\! $\Im(\tilde{\lambda}[k])\!\leq\! H_I\! +\! \epsilon$}
\State $\text{append}(\tilde{\lambda}^{\text{xtop}},\tilde{\lambda[k]})$
\EndIf
\If {\!$\Im(\tilde{\lambda}[k]\!\leq\! L_I\! +\! \epsilon$\! and\! $\Im(\tilde{\lambda}[k])\!\geq\! L_I\! -\! \epsilon$}
\State $\text{append}(\tilde{\lambda}^{\text{xbot}},\tilde{\lambda[k]})$
\EndIf
\EndIf
\EndFor
\State sortAscendingByRealPart$(\tilde{\lambda}^{\text{xtop}})$
\State sortDescendingByRealPart$(\tilde{\lambda}^{\text{xbot}})$
\State $\text{cycle} \gets [L_R+iH_I]$ \Comment{Top left corner}
\For{$k = 1,\, \ldots,$ size($\tilde{\lambda}^{\text{xtop}}$)}
	\State $R \gets \Re(\tilde{\lambda}^{\text{xtop}}[k])$
	\State $I \gets \Im(\tilde{\lambda}^{\text{xtop}}[k])-\epsilon$
	\State $\text{append}(\text{cycle}, [R-\epsilon+iH_I,\, R-\epsilon+iI])$
	\State $\text{append}(\text{cycle}, [R+\epsilon+iI,\, R+\epsilon+iH_I])$
\EndFor
\State$\text{append}(\text{cycle}, [H_R+iH_I,H_R+iL_I])$ 
\For{$k = 1,\, \ldots,$ size($\tilde{\lambda}^{\text{xbot}}$)}
	\State $R \gets \Re(\tilde{\lambda}^{\text{xbot}}[k])$
	\State $I \gets \Im(\tilde{\lambda}^{\text{xbot}}[k])+\epsilon$
	\State $\text{append}(\text{cycle}, [R+\epsilon+iL_I,\, R+\epsilon+iI])$
	\State $\text{append}(\text{cycle}, [R-\epsilon+iI,\, R-\epsilon+iL_I])$
\EndFor
\State $\text{append}(\text{cycle}, [L_R+iL_I,\, L_R+iH_I])$ \Comment{Close cycle}
\State \Return $\text{cycle}$
\EndProcedure
\end{algorithmic}
\end{algorithm}

Some of the integrals require a path from $\infty^-$ to $\infty^+$. It must not cross any other cycle in the homology basis (continuous deformations that do not cross branch points are allowed). A path with a single sheet change is obtained by encircling a single branch point once. Since $\lambda_{2g+2}$ does not participate in any other cycle by construction, a possible path is constructed by the following procedure: Start with a square centered around the point $\lambda_{2g+2}$ with edges of length $2\epsilon$. Extend this to the real axis at $M =\max_k \Re(\lambda_k) + \epsilon$ by first moving in the positive direction parallel to the real axis and then in the imaginary direction towards the real axis. Finally include the real axis from $M$ to $\infty$. 
The pseudocode for the algorithm to obtain the path up to the point $M$ is provided in Algorithm~\ref{alg:path_to_infinity}. The obtained path is shown in the right panel of Fig.~\ref{fig:cycle_algo}.

\begin{algorithm}
\caption{Construction of path from $M^-$ to $M^+$}\label{alg:path_to_infinity}
\begin{algorithmic}[1]
\Procedure{PathToM}{$\tilde{\lambda}$}
\State $\epsilon \gets \frac{1}{2}\min_{k,l}\max (\Re(\tilde{\lambda}[k]-\tilde{\lambda}[l]), \Im(\tilde{\lambda}[k]-\tilde{\lambda}[l]))$
\State $M \gets \max_k \Re(\tilde{\lambda}[k])+ \epsilon$
\State $s \gets \text{sign}(\Im(\tilde{\lambda}[2g+2]))$
\State $R \gets \Re(\tilde{\lambda}[2g+2])$ 
\State $I \gets \Im(\tilde{\lambda}[2g+2])$ 
\State $\text{path} \gets [M, M+i(I+s\epsilon),R-\epsilon +i(I+s\epsilon)]$
\State$\text{append}(\text{path},[R-\epsilon +i(I-s\epsilon),R+\epsilon +i(I-s\epsilon)])$
\State$\text{append}(\text{path},[R+\epsilon +i(I+s\epsilon), M+i(I+s\epsilon), M])$
\State \Return $\text{path}$
\EndProcedure
\end{algorithmic}
\end{algorithm}

\subsection{Sheet changes and direction of cycles}
\label{sec:sheetchanges}

Each of the paths we just computed is determined by the $\lambda$-coordinate. To make sure the $P$-coordinate in~\eqref{eq:riemann_surface} (see also~\eqref{eq:defineP}), on which the differentials also depend, is continuous, we must find the places on each path where a sheet change takes place.
The position of the sheet changes depends on the specific implementation of the square root. As an example, the conventional definition of the square root function always returns a number with positive real part. Therefore $\sigma$ changes sign on the negative real line, that is, when $\Im(P^2) = 0$ and $\Re(P^2)<0$. We find these points by a standard root-finding algorithm for the function $\Im(P^2)$ on the straight line (cycle segment) represented by its endpoints: $[\nu_k,\nu_{k+1}]$. This returns a list of points, $\tilde{z}_{k,1},\ldots,\tilde{z}_{k,\tilde{N}_k}$, at which $\Im(P^2)=0$. This list is filtered to ensure $\Re(P^2)<0$ and that the path crosses a sheet boundary, yielding a list of $N_k$ sheet changes $z_{k,1},\ldots,z_{k,N_k}$ on segment $k$.

To determine the direction of the cycles and obtain a canonical basis, for each $j$ pick a point $\lambda$ that is common to $a_j$ and $b_j$. If $a_j$ and $b_j$ do not cross in $\lambda$ because they lie on different sheets, reverse all sheet signs of $b_j$. Finally define the direction of $b_j$ such that the intersection number $a_j\circ b_j$~\eqref{eq:canonicalbasis} equals $+1$.
See Fig.~\ref{fig:cycle_algo} for an example of a canonical basis showing sheet changes and direction of cycles. 

\subsection{Numerical evaluation of integrals over Riemann surface}
\label{sec:numint}

All integrals over the Riemann surface that arise in the inverse PNFT can be written in the form $\int_\gamma f(\lambda)/P(\lambda)\, d\lambda$, where $\gamma$ is a (not necessarily closed) path on the Riemann surface and $f(\lambda)$ is a function of the $\lambda$-coordinate only. The numerator also depends on the sheet index through $P(\lambda)$, see Eq.~\eqref{eq:defineP}.
To compute these integrals, we first divide the integration path into $N-1$ straight line segments represented by their respective endpoints, $[\nu_k, \nu_{k+1}]$, $k=1,\ldots,N-1$. On segment $k$, we have the $N_k$ sheet changes $z_{k,1},\ldots,z_{k,N_k}$, which further divides the path into segments between sheet changes.
The integral over segment $k$ is then given by a sum of integrals over the set of segments 
$[\nu_k,z_{k,1}]$, $[z_{k,1},z_{k,2}]$, $\ldots$, $[z_{k,N_k},\nu_{k+1}]$. To account for the sheet changes, each integral is multiplied by the respective sheet sign $\sigma$.
Defining $z_{k,0}\equiv \nu_k$ and $z_{k,N_k + 1}\equiv\nu_{k+1}$ for notational convenience, the integral over a path $\gamma$ can be expressed as:
\begin{equation}
\int_{\gamma}\frac{f(\lambda)}{P(\lambda)}d\lambda = \sum_{k=1}^{N-1}\sum_{l=0}^{N_k}(-1)^{(\sum_{m=1}^{k-1}N_m) +l}\int_{z_{k,l}}^{z_{k,l+1}} \frac{f(\lambda)}{\sigma P(\lambda)} d\lambda,
\label{eq:numericpathint}
\end{equation}
where the sum over $k$ sums the contributions from the $N-1$ segments of the path. The sum $\sum_{m=1}^{k-1}N_m$ in the exponent keeps track of the number of sheet changes that have occurred on the previous line segments.
Here $f(\lambda)$ and $\sigma P(\lambda)$ are functions of $\lambda$ only, see~\eqref{eq:defineP}. The integrand has singularities at the branch points since $P(\lambda_j)=0$. The integration paths however never cross a branch point, see Sec.~\ref{sec:homology_basis} and Appendix~\ref{app:math}.
The resulting integrals over the complex plane can therefore be evaluated by standard numerical methods.
We can now compute $\bm{A}, \bm{B}$, and $\underline{\delta}^+-\underline{\delta}^-$. From these $\bm{\tau}$, $\underline{k}$ and $\underline{\omega}$ are obtained straightforwardly.

\subsection{Evaluation of $|K_0|$, $\omega_0$ and $k_0$}
\label{sec:evaluatek0}

This leaves the computation of $|K_0|$, $\omega_0$ and $k_0$, which separates into two parts: First we compute the unique normalized differentials $d\Omega_0$, $d\Omega_1$ and $d\Omega_2$ which have their asymptotic behavior toward $\infty^\pm$ determined by Eqs.~\eqref{eq:OmegaAsymptotics}. From these differentials we isolate their subleading behavior by removing the divergence towards $\infty^\pm$.
The differentials defined by~\cite[below Eq. 4.3.5]{Belokolos1994}:
\begin{align}
d\widetilde{\Omega}_0 &= \frac{\lambda^{g}}{P(\lambda)}d\lambda,\label{eq:omeganorm0}\\ 
d\widetilde{\Omega}_1 &= \frac{\lambda^{g+1}-d_1 \lambda^g}{P(\lambda)}d\lambda,
\label{eq:omeganorm1}\\ 
d\widetilde{\Omega}_2 &= 4\frac{\lambda^{g+2}-d_1\lambda^{g+1}-d_2\lambda^g}{P(\lambda)}d\lambda,\\
d_1 &= \frac{1}{2}\left(\sum_{j=1}^{2g+2}\lambda_j\right)\text{,\,\,}
d_2 =\frac{1}{8}\left(\sum_{j=1}^{2g+2}\lambda_j^2 -2\sum_{j=1}^{2g+2}\sum_{k<j}\lambda_j\lambda_k\right),\label{eq:omeganorm2}\nonumber
\end{align}
have the proper leading divergence, with all subleading divergences removed. This can be shown by computing the Laurent series for these differentials near $\infty^\pm$. To normalize these differentials and make the integral over each $a_j$-cycle zero, we subtract a multiple of the normalized holomorphic differentials $d\psi_j$, Eq.~\eqref{eq:normdiff}, as they obey $\int_{a_j}d\psi_k = \delta_{j,k}$:
\begin{equation}
d\Omega_k = d\widetilde{\Omega}_k - \sum_{j=1}^g d\psi_j \int_{a_j}d\widetilde{\Omega}_k.
\label{eq:omeganorm}
\end{equation}

To evaluate the formal expressions~\eqref{eq:evalomega0} through~\eqref{eq:evalomega2}, we separate the path to and from $\infty^{\pm}$ into a path connecting $M^\pm$ on the two different sheets of the Riemann surface, and the real line $[M,\infty]$ (see right panel of Fig.~\ref{fig:cycle_algo}):
\begin{align}
-\log(-\frac{1}{4}|K_0|^2) &=\!\! \int_{M^-}^{M^+}\!\!\!\!\! d\Omega_0 -2\!\!\int_{1}^M\!\!\! \frac{d\lambda}{\lambda} +2\!\!\int_M^\infty\!\! (d\Omega_0\!-\!\frac{1}{\lambda}d\lambda),
\nonumber\\
\omega_0 &=\!\! \int_{M^-}^{M^+}\!\!\!\!\! d\Omega_1 -2\!\!\int_{0}^M\!\!\! d\lambda +2\!\!\int_M^\infty\!\! (d\Omega_1\!-\! d\lambda),
\nonumber\\
k_0 &=\!\! \int_{M^-}^{M^+}\!\!\!\!\! d\Omega_2 -2\!\!\int_{0}^M\!\!\! 4\lambda d\lambda +2\!\!\int_M^\infty\!\! (d\Omega_2 \!-\! 4\lambda d\lambda).
\label{eq:k0num}
\end{align}
The integrals from $M^-$ to $M^{+}$ on the right-hand side contain a sheet change. We evaluate them by bringing the integrands $d\Omega_k$ into the form $[f(\lambda)/P(\lambda)]d\lambda$ as in~\eqref{eq:numericpathint}. Using the definitions of the (normalized) differentials~\eqref{eq:holomorphic_differentials} and~\eqref{eq:normdiff} in~\eqref{eq:omeganorm} together with~\eqref{eq:omeganorm0} we have, for example, for the integral over $d\Omega_0$,
\begin{equation}
f(\lambda) = \lambda^g - \sum_{j=1}^g \sum_{k=1}^g \alpha_j(\bm{A}^{-1})_{j,k}\lambda^{k-1}
\end{equation}
and similarly for the other integrals.
Here we have defined the normalization constants
\begin{equation}
\alpha_j := \int_{a_j}d\widetilde{\Omega}_0,
\end{equation}
for which in turn $f(\lambda)=\lambda^g$. 
The second integrals on the right-hand sides of~\eqref{eq:k0num} can be evaluated analytically. The last ones do not contain any sheet changes since $\lambda\in\mathds{R}$. They are guaranteed to be finite and can be evaluated using standard integration routines for improper integrals.

\subsection{Computing the inverse PNFT}
\label{sec:pnft_algorithm}

The computation of the inverse PNFT now
proceeds as follows. Obtain the integration paths as described in Algorithms~\ref{alg:homology} to~\ref{alg:path_to_infinity}.
Determine the sheet changes and the direction of the cycles in the homology basis as described in Sec.~\ref{sec:sheetchanges}.
Numerically compute integrals over the cycles as described in Sec.~\ref{sec:numint}. In particular, compute $\underline{\omega}$~\eqref{eq:periods}, $\underline{k}$~\eqref{eq:periods2}, $\bm{A}$~\eqref{eq:Amat} and the period matrix $\bm{\tau}$~\eqref{eq:tau}. Let $\underline{\delta}^-=0$ and compute $\underline{\delta}^+$ from the phase difference $\underline{\delta}^+-\underline{\delta}^-$~\eqref{eq:delta}. Compute $|K_0|$, $\omega_0$ and $k_0$ as described in Sec.~\eqref{sec:evaluatek0}.
Finally insert these constants into the solution~\eqref{eq:thetasol} and evaluate the expression for given $t$ and $z$.

\subsection{Numerical examples}
\label{sec:numerical_examples}

\begin{figure}[t]
\begin{center}
 \includegraphics[width=0.4\textwidth]{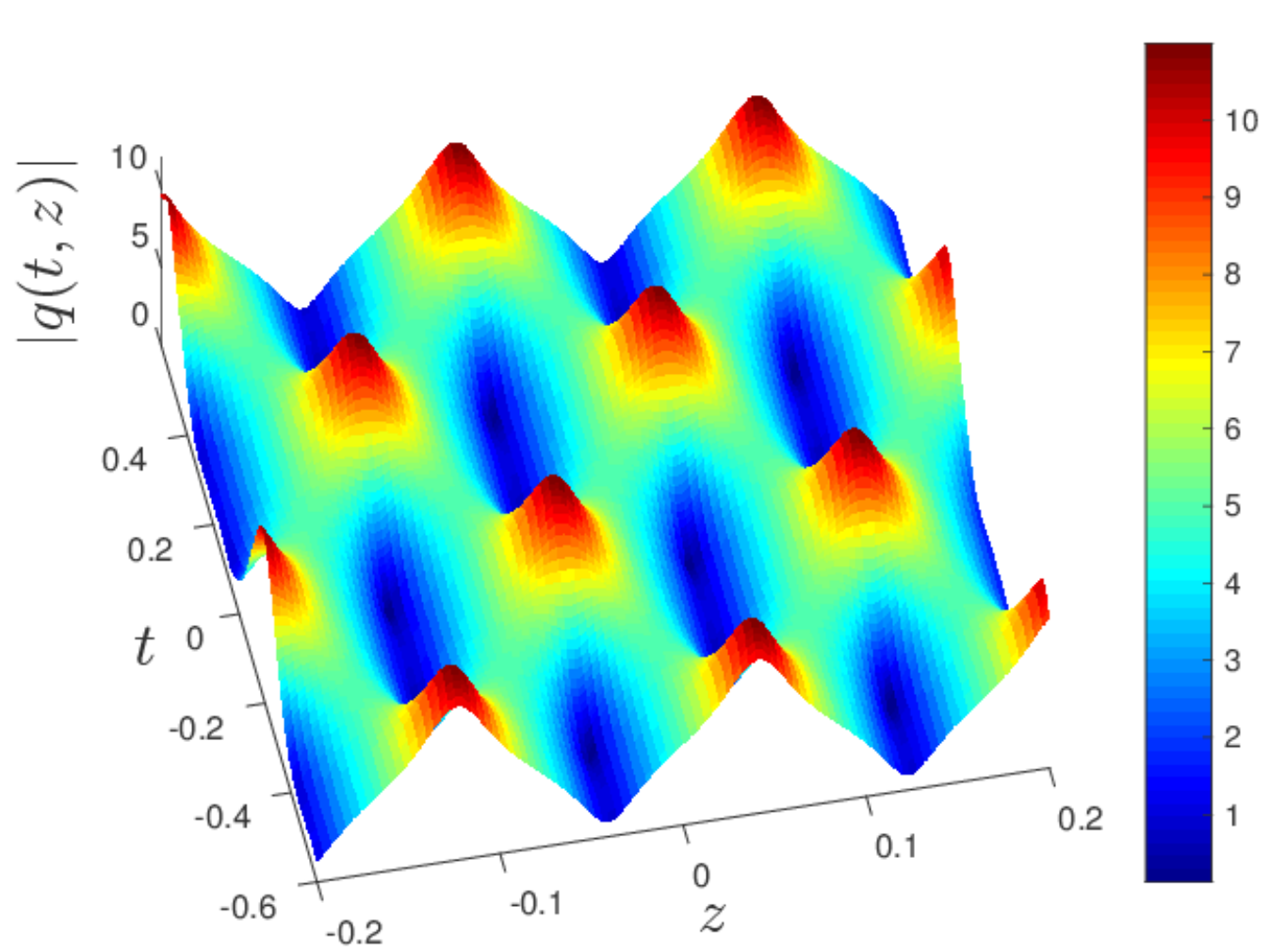}
\end{center}
 \caption{\label{fig:numerical_example_1} Amplitude of a genus-2 solution with spectrum $\{\lambda_j\}=\{-a\pm bi,\pm ci, a\pm bi\}$, with $a=1$, $b=3$ and $c=5$. The amplitude is periodic in time and space.}
\end{figure}

We provide two test cases for our algorithm. The first reproduces a result from Ref.~\cite[Fig. 5]{Smirnov2013} for a genus-2 solution. The spectrum has the symmetric form $\{\lambda_j\}=\{-a\pm bi,\pm ci, a\pm bi\}$, with $a=1$, $b=3$ and $c=5$. The associated algebraic curve in~\eqref{eq:riemann_surface} can be expressed in the form $P^2 = (\lambda^2 + c^2)(\lambda^4 + 2(b^2 - a^2)\lambda^2 +(a^2+b^2))$. In addition to complex conjugation, the spectrum has an additional involution, namely negation: only squares of $\lambda$ appear in the definition of the curve. As a consequence, there exist double coverings of two genus-1 surfaces by the genus-2 surface~\cite{Belokolos1994}, which allow algebraic reduction: The cycles on the genus-2 surface can be expressed in terms of cycles on two genus-1 surfaces with integer coefficients. 
Consequently the parameters can be expressed in terms of Jacobi elliptic functions instead of hyperelliptic ones~\cite{Smirnov2013}. Frequencies and wavevectors become integer multiples of each other: $\omega_1=0$, $\omega_2=\omega$ and $k_1=2k_2$. Since $\omega_0=0$ the solution is periodic in time.
Here we compute the waveform with our general algorithm. The parameters of the theta function representation \eqref{eq:thetasol} are summarized in Table~\ref{tab:parameters_1}. The amplitude of the waveform is shown in Fig.~\ref{fig:numerical_example_1}.
As one can see from the table, the algorithm reproduces the properties that follow from the symmetry of the spectrum with high accuracy.

The second example is a genus-3 solution, which corresponds to one of the waveforms of the nonlinear frequency amplitude modulation scheme introduced in Sec.~\ref{sec:applications} below. The spectrum is given by $g_1=g_4=5$ and $g_2=g_3=7$ in Eq.~\eqref{eq:nfam_symbols}. By construction, the waveform is approximately time-periodic. Table~\ref{tab:parameters_2} indeed shows that the components of the $\underline{\omega}$-vector are approximate integer multiples of 40.

\begin{figure}[t]
\begin{center}
 \includegraphics[width=0.4\textwidth]{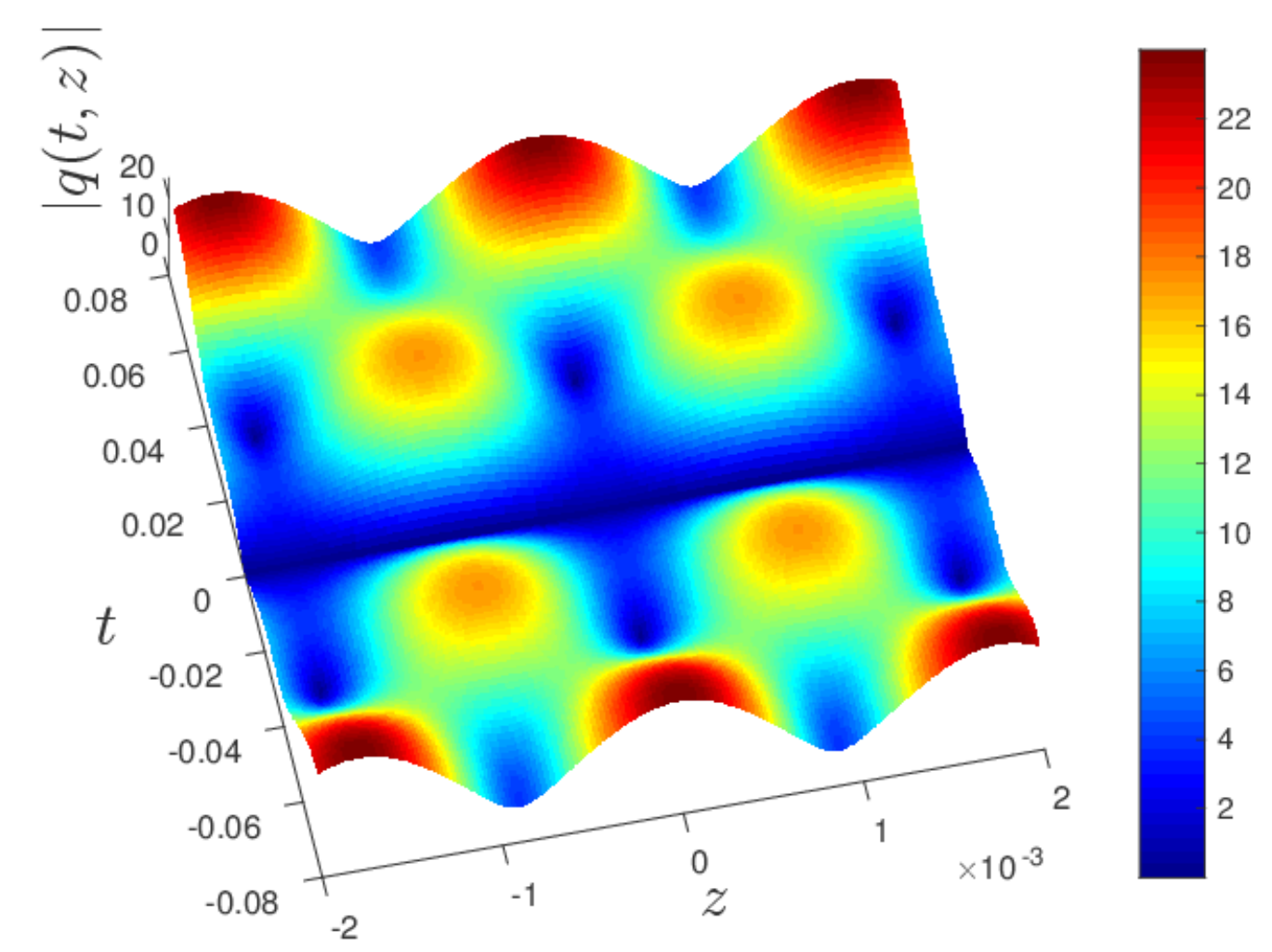}
\end{center}
 \caption{\label{fig:numerical_example_2} Amplitude of an approximately time-periodic solution. The solution corresponds to one out of the 256 signals in the nonlinear frequency amplitude modulation scheme described in Sec.~\ref{sec:nfam}.}
\end{figure}

\begin{table}[tb]
  \renewcommand{\arraystretch}{1.3}
	\caption{Numerical values of the nonlinear spectrum and theta function parameters of the waveform shown in Fig.~\ref{fig:numerical_example_1}. Imaginary parts in $\omega_0$ and $\underline{\omega}$ of the order of $e-16$ have been set to zero.}
	\label{tab:parameters_1}
	\centering
		\begin{tabular}{c|c}
		  \hline
			Spectrum & $\{-1\pm 3i,\pm 5i, 1\pm 3i\}$\\
			\hline
			$|K_0|$ & 3.6061\\
			$k_0$ & 78.8096\\
			$\underline{k}$ & [-76.3942, 38.1971]\\
			$\omega_0$ & 0\\
			$\underline{\omega}$ & [0, 8.4308]\\
			$\underline{\delta}^-$ & [-1.9104i, 3.1416\,+\,0.9552i]\\
			$\bm{\tau}$ & $\begin{pmatrix}
				0.7567i & -0.5 -0.3783i\\
				-0.5 - 0.3783i & 0.6918i\\
			\end{pmatrix}$\\
			\hline
		\end{tabular}
\end{table}

\begin{table}[tb]
  \renewcommand{\arraystretch}{1.3}
  \caption{Numerical values of the nonlinear spectrum and theta function parameters of the waveform shown in Fig.~\ref{fig:numerical_example_2}. Imaginary parts in $\underline{k}$ of the order of $e-13$ have been set to zero.}
	\label{tab:parameters_2}
	\centering
		\begin{tabular}{c|c}
		  \hline
			Spectrum & $\{-30\pm 5i,-10\pm 7i,10\pm 7i,30\pm 5i\}$\\
			\hline
			$|K_0|$ & 6.4261i\\
			$k_0$ & 2.6069\\
			$\underline{k}$ & [0, -3.5445e+3, 0]\\
      $\underline{\omega_0}$ & -21.5630\\
			$\underline{\omega}$ & [-40.9973, -1.0643, -43.1260]\\
			$\underline{\delta}^-$ & [-5.9443i, 7.3887i, 3.1416\,+\,3.7084i]\\
			$\bm{\tau}$ & $\begin{pmatrix}
				4.9497i & -0.5 - 3.4209i & -1.8921i\\
				-0.5-3.4209i & 3.1654 & 0.5+1.5363i\\
				-1.8921i & 0.5+1.5363i & 1.1804i
			\end{pmatrix}$\\
			\hline
		\end{tabular}
\end{table}

\section{Applications}
\label{sec:applications}

\subsection{Nonlinear frequency amplitude modulation}
\label{sec:nfam}

Although all finite-gap waveforms are solutions to the NLSE, they are time-periodic only when all frequencies $\omega_j$, $j=0,\ldots,g$ are commensurate. This is not the case in general. When all frequencies except $\omega_0$ are commensurate, the solution is time-periodic up to a phase. We refer to such a solution as quasi-periodic. 
In order to recover the spectrum at the receiver via the forward PNFT-algorithm~\cite{Wahls2015} it suffices to construct quasi-periodic solutions (see Part II of the paper for details).

One method to obtain a set of quasi-periodic solutions from finite-gap solutions is to start from
the nonlinear spectrum for a plane-wave solution $q(t) \equiv K_0$:
\begin{equation}
\lambda^\pm_k = \pm\sqrt{\left(\frac{\pi k}{T_0}\right)^2-|K_0|^2}, k\in\mathds{Z}.
\end{equation}
When $K_0=0$, all points in this spectrum become double points on the real axis. These double points do not provide any dynamics to $q$~\cite{Tracy1984}. However, a perturbation that splits a double point will generate an approximately quasi-periodic signal with the period $T_0$. In the low power limit the real part of each pair of perturbed double points can be interpreted as determining the frequency, while the imaginary part determines the spectral amplitude of that component. This suggests that one can create an approximately quasi-periodic solution by taking the real parts of all eigenvalues equidistant, while modulating their imaginary part. 
Taking the same real parts in the set of spectra that forms the constellation leads to approximately the same quasi-period of the associated waveforms.
We call this proposal nonlinear frequency amplitude modulation (NFAM). As a specific example we show $4\times4$ - NFAM. We modulate 4 different nonlinear frequencies independently at 4 possible levels, corresponding to $4^4=256$ signals.

We modulate the nonlinear spectrum by generating $4\times 2$ bits per symbol. Each set of two bits is Gray-mapped to the set of possible $g_j\in\{5,7,9,11\}$. The resulting 4 integers determine one symbol with main spectrum:
\begin{equation}
s_{g_1,g_2,g_3,g_4} = \{-30+g_1i, -10+g_2i, 10+g_3i, 30+g_4i\}.
\label{eq:nfam_symbols}
\end{equation}
For each symbol, the auxiliary spectrum is fixed by letting $\underline{\delta}^+=0$.
Note that the four $g_j$ are completely independent.

\begin{figure}
 \center{\includegraphics[width=0.5\textwidth]{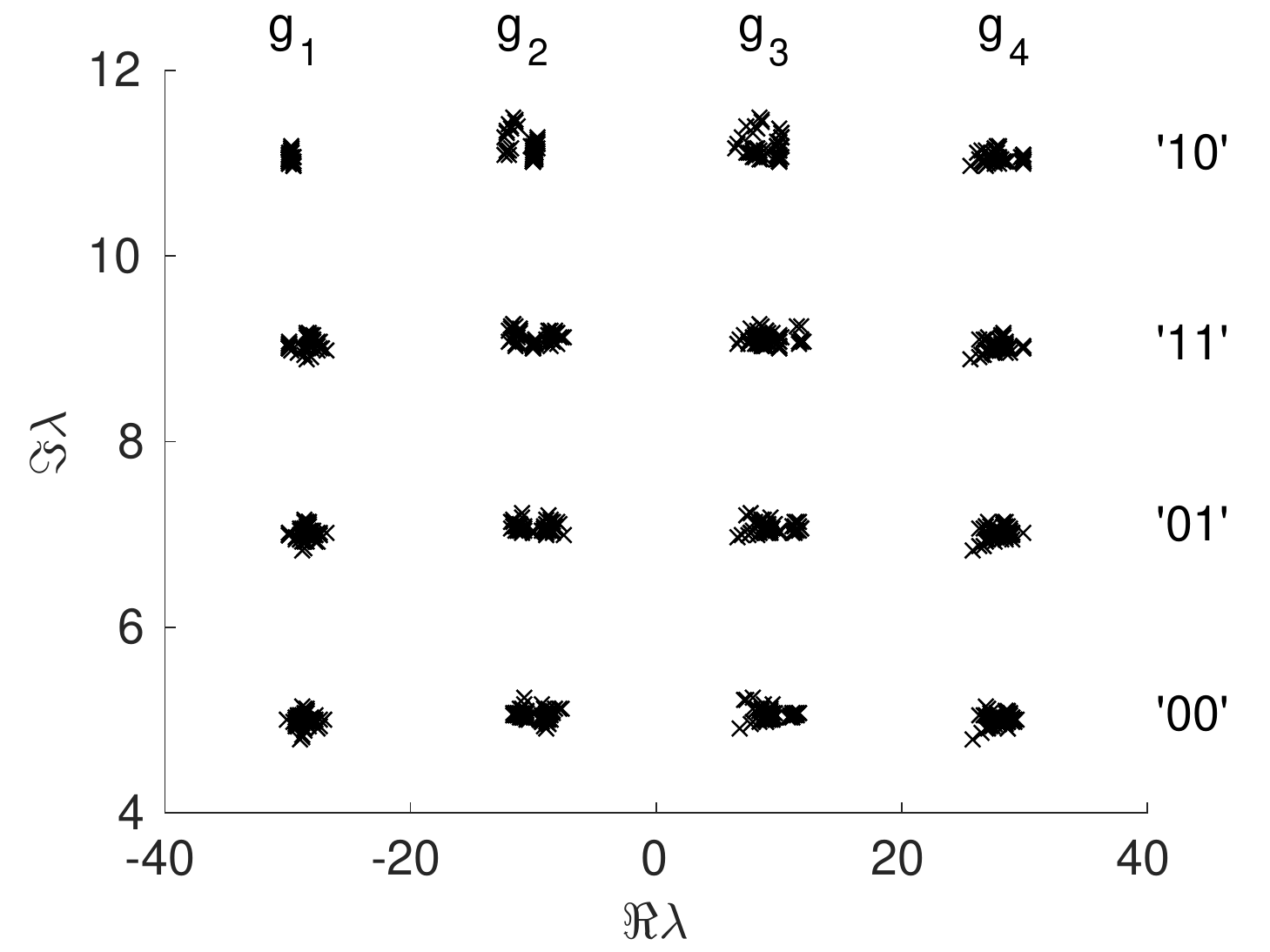}}
 \caption{\label{fig:lambda_error} The spectra of the periodic $4\times4$ NFAM waveforms. Each spectrum has one point in each column.  The total constellation allows the modulation of $4\times 2$ bits per symbol. These are obtained in back-to-back, without adding noise. The constellation is slightly irregular due to period matching of the waveforms.}
\end{figure}

The inclusion of multiple points in the nonlinear spectrum changes the differentials $dU_j$. Therefore the frequencies are no longer exactly commensurate and the waveforms obtained by this construction are not exactly quasi-periodic. In the limit that all imaginary parts go to 0, the $\omega_j$ become multiples of 40.
We force all 256 signals to become exactly quasi-periodic with the same period by setting the values of $\omega_i$ manually to multiples of 40. We then compute the initial condition for each of the spectra by evaluating the theta function ratio with the new $\underline{\omega}$-vector. The adjustment of $\underline{\omega}$ changes the nonlinear spectrum. We therefore compute the resulting spectra numerically using an algorithm for the forward transform. They are shown in Fig.~\ref{fig:lambda_error}. 
Each of the 256 spectra is affected differently by the change of $\omega$, resulting in a small spread around the original points. Hence they are no longer exactly given by~\eqref{eq:nfam_symbols}.
Note that this periodization scheme cannot be applied when taking the Riemann-Hilbert problem approach to the inverse PNFT, since in that case the analytical parametrization of $q(t,z)$ is not available.

\begin{figure}[h]
 \center{\includegraphics[width=0.5\textwidth]{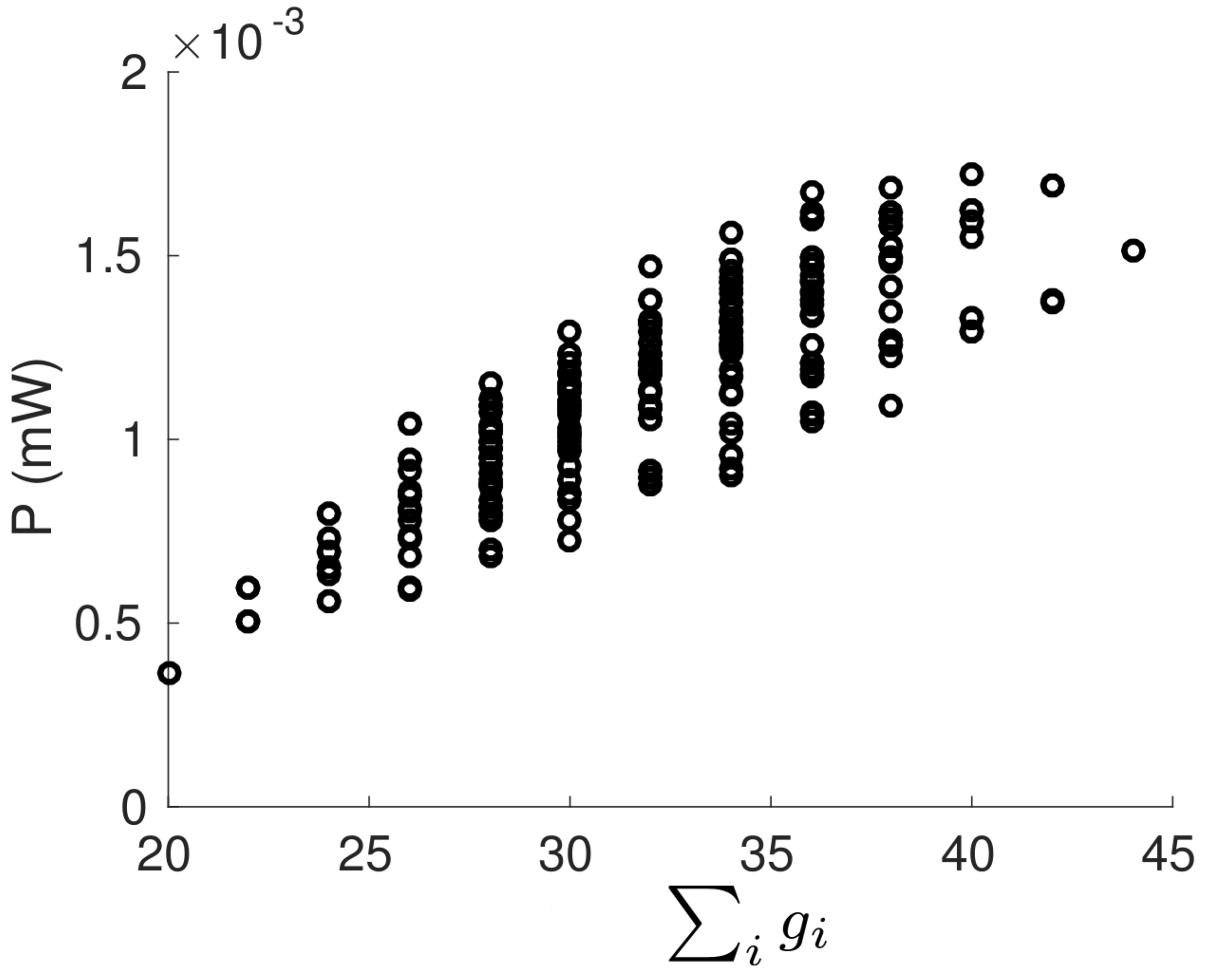}}
 \caption{\label{fig:power} The power of the  256 different NFAM signals, as a function of the sum of the imaginary parts of the spectrum, $\sum_i g_i$.  This plot suggests that the power of a waveform increases on average when the sum of the imaginary part of the main spectrum increases.
}
\end{figure}

When designed with a quasi-period of $0.5$ ns, the modulated frequencies correspond to $\pm 1$ and $\pm 3$ GHz. For the parameters of standard single mode fiber (see Table~\ref{tab:fiber_parameters}), the lowest power symbol has $P =-4.4$ dBm, while the highest power symbol has $P = 2.5$ dBm. The distribution of the power of all symbols can be seen in Fig.~\ref{fig:power}. The power increases as $\sum_i g_i$ increases, providing a heuristic for power control in the nonlinear domain. Note that the linear trend that is observed cannot be valid for all values of $\sum_i g_i$ as the trend line does not pass the origin.

\begin{table}[tb]
  \renewcommand{\arraystretch}{1.3}
  \caption{Fiber parameters used in the simulations.}
	\label{tab:fiber_parameters}
	\centering
\begin{tabular}{c|c}
\hline
Parameter & Value\\
\hline
$\alpha $    & 0.2 dB/km    \\
$\beta_2$ & -21.5 ps$^2$ km$^{-1}$\\
$\gamma $    & 1.3 W$^{-1}$ km$^{-1}$\\
$L_{span}$& 75 km\\
\hline  
\end{tabular}
\end{table}

Based on this constellation, we simulate the transmission of $10^5$ symbols, corresponding to $8\cdot10^5$ bits. We apply the split-step Fourier method to simulate the optical channel including attenuation. The setup we simulate consists of 75 km spans, with noise-loading between spans corresponding to a noise figure of 5.5 dB. To account for the attenuation we use the transformed lossless model~\cite{Kamalian2017}, with an effective nonlinearity parameter  $\gamma_\text{eff} = \gamma[1-\exp(-\alpha L_{\text{span}})]/\alpha$, where 
$\alpha$ is the attenuation parameter, $\gamma$ is the nonlinearity parameter of the fiber and 
$L_{\text{span}}$ is the length of a fiber span. We estimate the necessary cyclic prefix to prevent inter-symbol interference by
\begin{equation}
\Delta T=2\pi|\beta_2|L B\label{eq:memory},
\end{equation}
where $L$ is the total propagation distance, $B=8$ GHz is the bandwidth of the signal, and $\beta_2$ is the group velocity dispersion parameter. We choose $\Delta T = 1.5$ ns (corresponding to 3 periods of cyclic prefix), which makes the predicted inter-symbol interference free transmission distance 1500 km.

At the receiver the cyclic prefix is removed, and the forward PNFT is computed by means of the FNFT-package~\cite{Wahls2018}. The obtained spectrum is reduced to the four spectral points with largest imaginary part. These are demapped individually, with the point with the smallest real part corresponding to $g_1$, etc. Note that the center frequency for different symbols could be chosen independently, because the detection does not depend on the position of the spectrum along the real axis. This could in principle be used to adjust the group velocity to reduce the necessary cyclic prefix length. An example for the adjustment of the group velocity is provided in Part II~\cite{GoossensII}.

\begin{figure}
 \center{\includegraphics[width=0.5\textwidth]{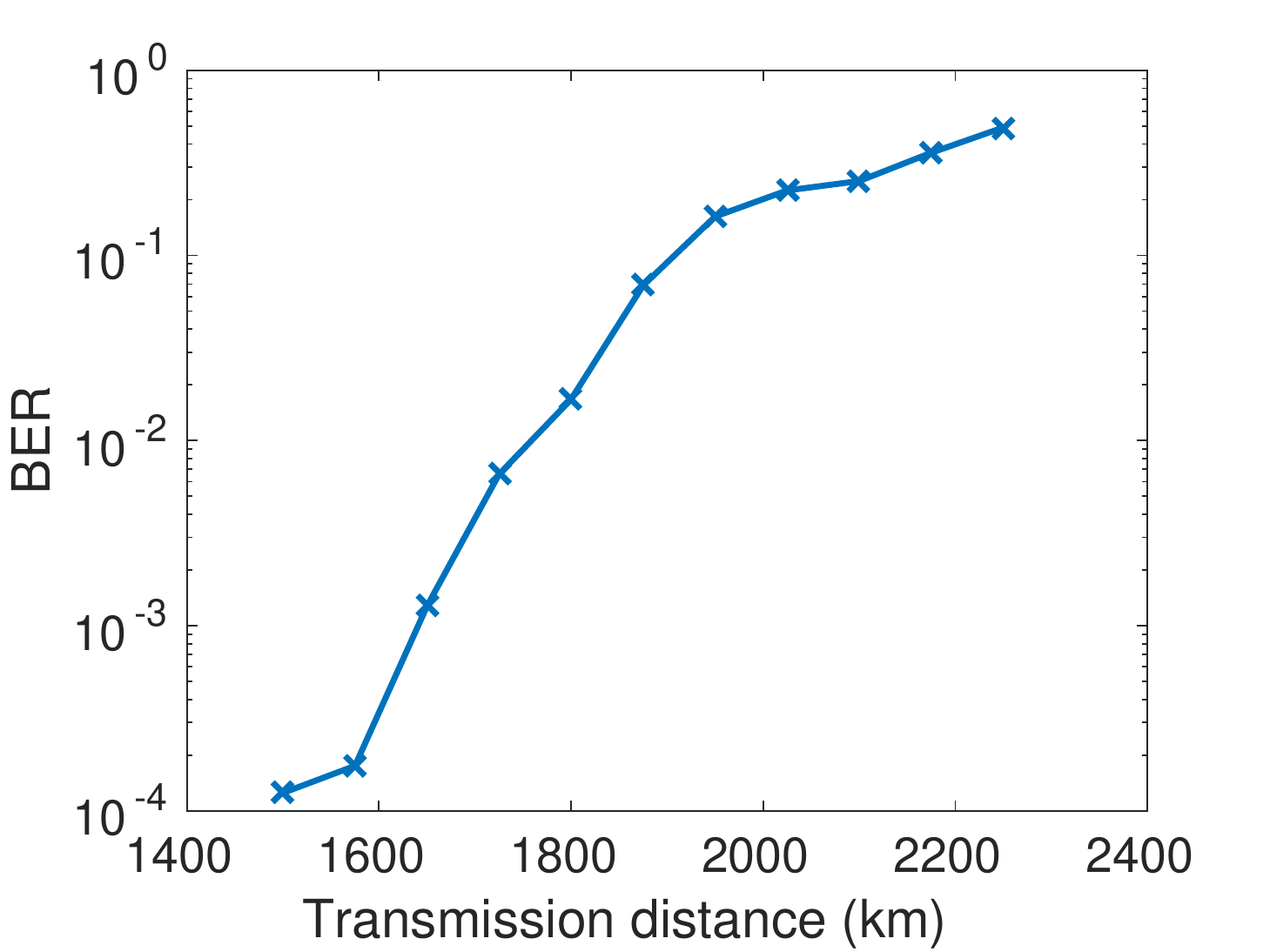}}
 \caption{\label{fig:BER-curve} Bit error rate as a function of distance for the $4\times$5 NFAM-scheme. The reach for a BER threshold of $10^{-3}$ is 1575 km, which is close to the estimate based on group velocity dispersion.}
\end{figure}

The resulting BER-curve is shown in Fig.~\ref{fig:BER-curve}. We observe a BER below $10^{-3}$ up to 1575 km. For larger distances the BER grows due to increasing inter symbol interference. The breakdown of this scheme occurs at a distance that roughly agrees with the onset of ISI according to Eq.~\eqref{eq:memory}. 
By artificially putting one period of the waveforms on a vanishing background and computing the forward NFT with vanishing boundary conditions, we can obtain a rough measure of their solitonic component (the energy contained in the discrete spectrum). In this case it is small.
In Part II~\cite{GoossensII} we show an example where waves with a large solitonic component lead to a increased transmission reach relative to the expectation provided by Eq.~\eqref{eq:memory}.
The spectral efficiency in this example is approximately 0.67 bits/s/Hz. While this is less than obtained in the scheme of Ref.~\cite{Kamalian2017b} (2.8 bits/s/Hz over 1000 km), here we provide the first scheme which simultaneously modulates more than two degrees of freedom.

Note that when increasing the power of the waveforms this modulation scheme will eventually break down, because the interaction between the eigenvalues changes the frequency vector too much. This would reduce the distance between symbols.

\subsection{Phase modulation}

The nonlinear phase vector $\underline{\delta}^+$ as defined in Eq.~\eqref{eq:deltapm} is constant during propagation, and choosing a vector $\underline{\delta}^+$ implicitly fixes the auxiliary spectrum. This suggests the modulation of the phase $\underline{\delta}^+$ to encode information on the auxiliary spectrum in the interval $[0,2\pi]$.
At the receiver, the forward NFT returns the main spectrum $\lambda_k$ and the auxiliary spectrum $\mu_j(t_0,L)$ where $L$ is the propagation distance, and $t_0$ is the time at the start of the processing frame for the forward NFT. This auxiliary spectrum can then be inserted into Eq.~\eqref{eq:deltapm} to obtain $\delta_j^\pm$ in terms of known parameters:
\begin{align}
\frac{1}{2\pi}\delta^\pm_j &= \int_{p_0}^{\infty^\pm}d\psi_j -\frac{1}{2}\tau_{jj} + \sum_{k=1}^g\int_{a_k}d\psi_k(p')\int_{p_0}^{p'}d\psi_j(p)\nonumber\\
&- \sum_{k=1}^g\int_{p_0}^{\mu_k(t_0,L)}d\psi_j-\frac{1}{2\pi}(\underline{\omega}t_0+\underline{k}L).
\end{align}
Note that all terms, except for the integrals to $\mu_k(t_0,L)$ in the second row are independent of the auxiliary spectrum, and can be precomputed. One extra necessary ingredient required for demodulation is an algorithm that can provide paths from the base-point $p_0$ to any point $\mu_k(t,z)$ in the auxiliary spectrum without crossing any of the $a$- or $b$-cycles. Alternatively, one could also use any path from $p_0$ to $\mu_k(t,z)$, and compensate for each crossing by subtracting the appropriate $a$- or $b$-cycles.  We currently consider both these approaches too complex for online computation and therefore do not develop the scheme further in this paper. 

\subsection{Genus modulation}

\begin{figure}
 \center{\includegraphics[width=0.5\textwidth]{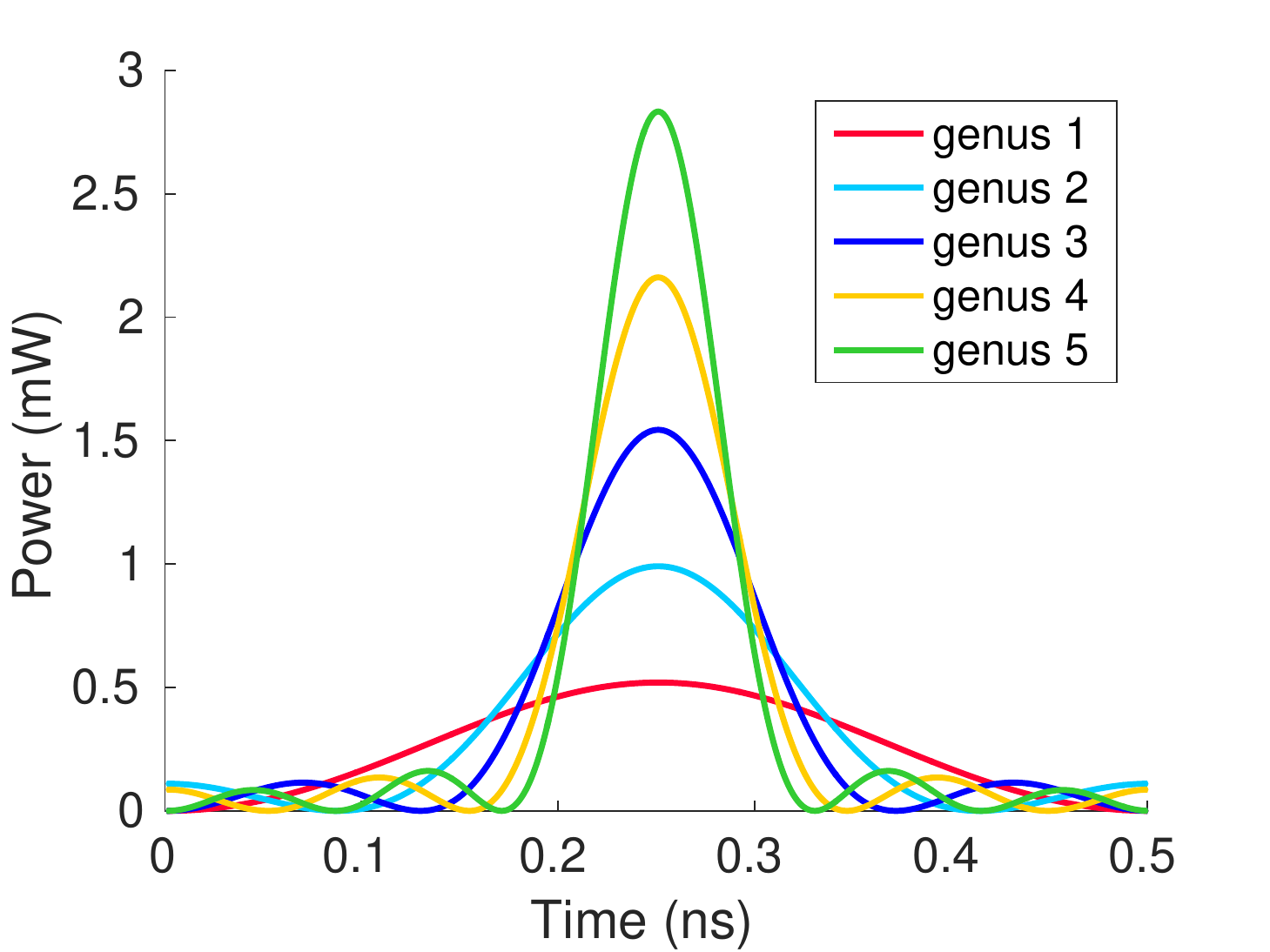}}
 \caption{\label{fig:Genusmodulation} Five waveforms corresponding to five different genera. The maximum power of the pulse increases with genus.}
\end{figure}

Another application for the inverse PNFT is an information transmission scheme in which information is encoded in the genus of the nonlinear spectrum. The advantage of such a scheme lies in the robustness of the genus to non-ideal channel properties. In the simplest scenario, the genus of the transmitted waveform is either $0$ or $1$. In that case, this scheme bears resemblance to soliton communication based on on-off-keying, with the difference that the genus-0 solution corresponds to a nonzero background.

Five example waveforms with genus 1 to 5 are shown in Fig.~\ref{fig:Genusmodulation}. Their spectra are formed by defining $\lambda_k = 20k+5i$ and taking the lowest $g+1$ eigenvalues $\lambda_k$ for the solution with genus $g$. As for NFAM, we ensure these solutions are periodic by adjusting the $\underline{\omega}$-vector. The genus determines the overall structure of the waveform,. It is therefore conceivable that noise cannot easily alter the solution of one genus to resemble that of another. The only way to generate new points in the spectrum is by splitting degenerate points. These are all located on the real axis. Therefore new eigenvalues can only appear close to the real axis, which allows them to be simply filtered out. In this sense, encoding information in the genus provides a ``topological protection'' of information, making the symbols extremely robust against adverse channel conditions. 

The power of the genus 5 symbol in Fig.~\ref{fig:Genusmodulation} is approximately twice as much as for the genus 1 symbol in the same figure. We observe that in this case the genus determines the number of local maxima in the periodic solution. This is similar to what one would observe when adding $g+1$ plane waves together, and can be understood by considering the beating of  the $g+1$ linear frequencies.

Because no information is encoded in the actual position of the points of the spectrum in the complex plane, several degrees of freedom are available to ensure the waveforms in the resulting constellation have similar group velocity and obey certain boundary conditions.

In Part II of this paper we show that this freedom can minimize nonlinear interference and reduce the necessary cyclic prefix. We also show that taking into account these constraints leads to irregular constellations, which complicates the demapping. To find the optimal modulation scheme is an optimization problem that falls outside the scope of the current text, and is left for future work.

\section{Conclusion}
\label{sec:conclusion}

In this paper we have formulated the exact inverse PNFT based on the algebro-geometric integration method in terms of the communication problem. We have provided an algorithm for the computation of the exact inverse PNFT, including a fully automated method to construct a homology basis. 

For increasing genus, the computational burden of the evaluation of the Riemann theta function grows exponentially, and becomes the bottleneck of the algorithm. While the Riemann-Hilbert approach of Ref.~\cite{Kamalian2018} is computationally less demanding, the inverse transform based on the algebro-geometric approach has the advantage that an analytical expression for the waveform is available. We have exploited this property to obtain waveforms that are quasi-periodic. Even if the proposed algorithm may not be practical for the online computation of waveforms at the transmitter, we demonstrate in Part II~\cite{GoossensII} that signals can be precomputed and stored in a look-up table for online signal generation. 

As applications of our algorithm, we have discussed three possible modulation schemes for PNFT based transmission. The first, which we refer to as nonlinear frequency amplitude modulation, modulates four degrees of freedom. This exceeds the number of degrees of freedom in previously proposed PNFT modulation schemes, and can be extended to more degrees of freedom. To achieve competitive spectral efficiency, it will be essential to modulate the auxiliary spectrum. Phase modulation, though theoretically appealing, requires construction of an integration path and numerical integration, which may be challenging to implement in online computation. Modulation of the genus has the advantage that it is highly robust against imperfect channel conditions. Since only the genus and not the precise spectrum need to be retrieved at the receiver, it may allow simplification of the forward transform.

We have seen that the design of practical modulation schemes is not straightforward, owing to the nonlinear character of the problem. In Part II~\cite{GoossensII} we investigate different aspects of the waveform design relevant for practical implementation in more detail. We develop a modulation scheme taking these constraints into account and assess its performance in a transmission experiment.

\section{Acknowledgments}

We  would like to thank Sander Wahls, Shrinivas Chimmalgi and Peter Prins for fruitful discussions. 

\appendix

In this Appendix we sketch the derivation of the analytical expression for finite-gap solutions of the focusing nonlinear Schr\"odinger equation (NLSE).
The aim is introduce the underlying mathematical notions, to outline the required steps in the derivation, to connect to the existing specialized literature and to provide further insight into the origin of the equations that enter the algorithm for the inverse periodic nonlinear Fourier transform.

Due to space constraints, it is not possible to provide a complete and mathematically rigorous exposition that contains all required mathematical proofs. Rather the aim here is to connect the different parts of the derivation that are spread out over the existing mathematical and physics literature. To this end, we have unified the notation and chose the convention relevant to the optical channel.

\subsection{Mathematical notions}
\label{app:math}

The treatment of the algebro-geometric approach to the inverse PNFT requires a number of mathematical notions which we introduce here for completeness. The goal is to provide an intuitive understanding of the underlying mathematics. A mathematically rigorous introduction of these notions is for example given in the book of Belokolos et al.~\cite{Belokolos1994}.
An excellent practical introduction to the topic of algebraic geometry and the theory of Riemann surfaces is given in Ref.~\cite{Bobenko2011}. 

We start with the basic notion of a Riemann surface. A Riemann surface 
is a one-dimensional complex manifold, which locally looks like a continuous deformation of the complex plane. Globally it can have a non-trivial topology, for example that of a torus.
Here topology or topological structure of a geometric object refers to properties which are preserved under \emph{continuous} deformations such as twisting, bending or stretching, but not under discontinuous ones such as tearing or gluing. For example, a coffee cup with one handle can continuously be deformed into a closed surface with one hole and is said to have the same topology as a torus. The two objects are topologically equivalent.

\begin{figure}[t]
\begin{center}
 \includegraphics[width=0.5\textwidth]{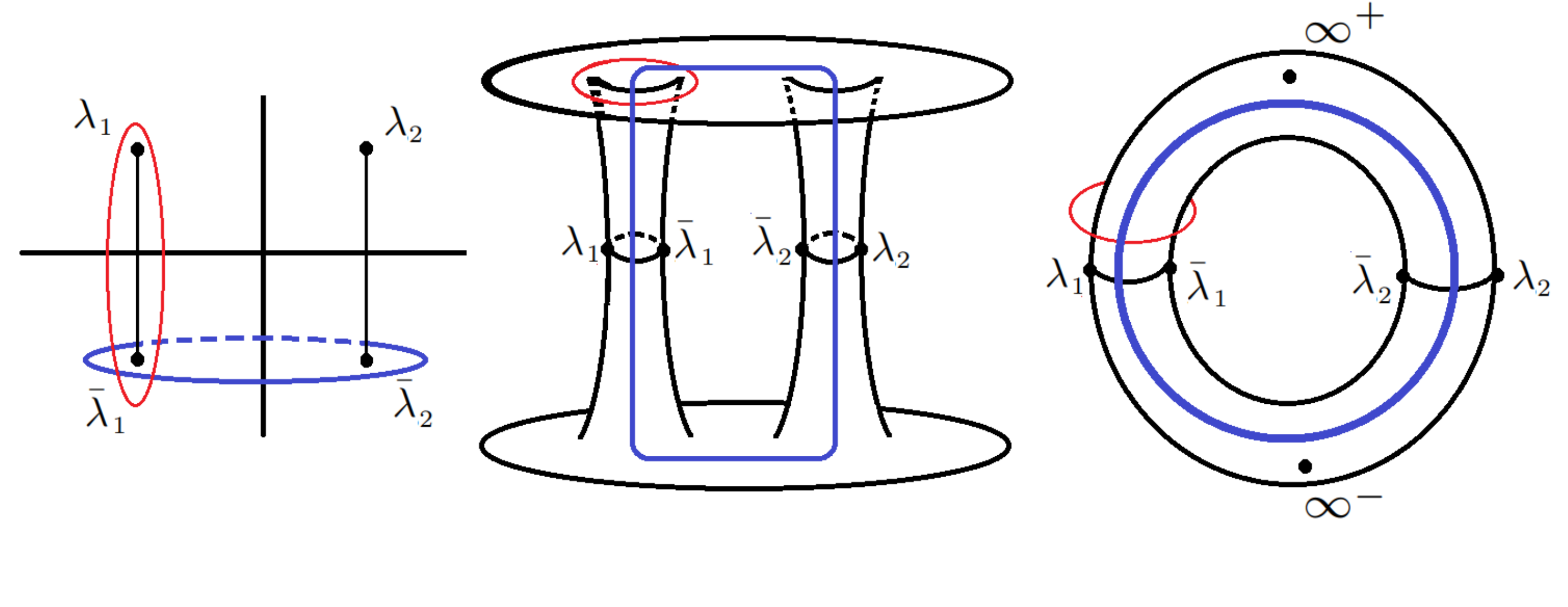}
\end{center}
 \caption{\label{fig:torus} 
Left: Illustration of the complex plane with branch cuts (thin vertical lines). Middle: two sheets of the complex plane glued together at the branch cuts. Right: Resulting toroidal topology of the Riemann surface. Note the sheet change (left and middle) in one of the cycles (blue, thick line).}
\end{figure}

Riemann surfaces are important in the study of the global behavior of holomorphic functions, that is, complex-valued functions which are continuously differentiable in the neighborhood of every point in their domain. 
They arose in the context of algebraic expressions defining multi-valued functions. 
Consider the relation $P^2=\lambda$. The function $P(\lambda) = \sqrt{\lambda}$ defined as the square root with positive real part necessarily is discontinuous across the negative real axis. 
It can be made continuous and even holomorphic by excluding the negative real axis from its domain.
Similarly, consider the (hyperelliptic) algebraic curve
\begin{equation}
\Gamma: \left\{(P,\lambda)\textrm{, }P^2 = \prod_{k=1}^{g+1} (\lambda-\lambda_k)(\lambda-\bar{\lambda}_k), P,\lambda\in\mathds{C}\right\},
\label{eq:app:riemann_surface}
\end{equation}
where $\lambda_1,\ldots,\lambda_{g+1}$ are complex numbers with positive imaginary part. The number $g$ is called the genus.
The Riemann surface of the multi-valued function
\begin{equation}
P(\lambda) = \sigma \sqrt{\prod_{k=1}^{g+1} (\lambda-\lambda_k)(\lambda-\bar{\lambda}_k)},
\label{eq:defineP}
\end{equation}
where $\sigma=\pm 1$, is the surface on which $P$ can be considered a single-valued analytic function. A point on it is denoted by $p=(P,\lambda)$. The surface is obtained by considering two sheets of the complex plane, one for each value of the sheet index $\sigma$. $P(\lambda)$ behaves as $\sqrt{\epsilon}$ times a holomorphic function in the vicinity $\lambda=\lambda_k+\epsilon$ of each of the $\lambda_k$ (and their conjugates). 
At these points $P=0$ and the sheets meet. They are called branch points. By fixing the definition of the square root, we obtain lines where $P(\lambda)$ changes sign. These are the $g+1$ branch cuts. Fig.~\ref{fig:torus} (left) shows their possible locations. 
Pulling out a tube from each of the branch cuts and gluing the tubes together (after flipping the lower sheet about the real axis) yields the configuration shown in Fig.~\ref{fig:torus} (middle).
Considering the projective plane $S=\mathds{C}\cup\{\infty\}$, the sheets can be deformed to closed spheres, connected by two tubes: the topological equivalent of a torus, Fig.~\ref{fig:torus} (right).
For details on this construction, we refer to~\cite{Teleman2003}.

The number of branch cuts or tubes is $g+1$ and the genus $g$ counts the number of topological holes. The topological structure of $\Gamma$ is hence completely determined by the number of roots $\lambda_k$.
For $g=0$, the surface is topologically equivalent to a sphere, for $g=1$ to a torus, for $g=2$ to a torus with 2 holes, and so on.
In case of the NLSE, the roots always appear in complex conjugate pairs~\cite{Belokolos1994}. As a consequence, there is always an even number of them. This implies in particular that $\lambda = \infty$ is not a branch point. Thus the two points $\infty^\pm$  corresponding to $\lambda = \infty$ are distinct.

We will further require the notion of Abelian differentials. 
In a neighborhood of any point on the Riemann surface, an Abelian differential can be written in the form $[f(\lambda)/P(\lambda)]d\lambda$. If $f/P$ is holomorphic, the differential is said to be holomorphic, or of the first kind.
When $f/P$ is meromorphic (holomorphic except for a set of isolated points which are the poles of $f/P$), the differential is called meromorphic. 

It is possible to define a basis of holomorphic differentials on a given Riemann surface. The number of elements in this basis equals the genus $g$~\cite[Sec. 2.4.2]{Belokolos1994}. Specifically, for the Riemann surface defined in~\eqref{eq:app:riemann_surface}, a basis of holomorphic differentials is given by
\begin{equation}
dU_j(p) = \frac{\lambda^{j-1}d\lambda}{P(\lambda)}, j=1,\ldots,g.
\label{eq:holomorphic_differentials}
\end{equation}
Note that, given the dependence of $P$ on $\lambda$, the differentials are no longer holomorphic for $j<1$ and $j>g$.
The formula
\begin{equation}
\Omega(p) = \int_{p_0}^p d\Omega(p')
\label{eq:abelian_integral}
\end{equation}
defines a multivalued function called an Abelian integral on the whole Riemann surface. Note that the integration over the path from $p_0$ to $p$ will in general involve sheet changes, which have to be taken into account. In the following and in accordance with the literature, we suppress the explicit dependence on $p$ if no ambiguity arises.

Abelian integrals can be divided into three distinct groups: those of the first kind locally correspond to holomorphic functions, and those of the second kind to meromorphic functions. All others correspond to the third kind and usually have logarithmic singularities.
The features of Abelian integrals of the first and second kind are completely described by their periods\footnote{Additional periods associated with the poles have to be invoked for Abelian integrals of the third kind~\cite[Eq. 2.4.6]{Belokolos1994}.}~\cite[Eq. 2.4.5]{Belokolos1994}.
To specify these periods on a surface of genus $g$, we can define a set of $2g$ directed distinct closed curves $a_1,\ldots,a_g, b_1,\ldots,b_g$, which can neither be deformed into each other nor shrunk to zero~\cite[Sec. 1.3.2]{Bobenko2011}. A closed curve is determined by specifying which branch points are inside it. An integral over the curve does not change its value under continuous deformations of the curve, which does not cross any branch points. I.e., the topology remains the same.
Any closed curve on the Riemann surface can be obtained as a linear combination of these basis elements. 

The intersection number $(a\circ b)_p$ of two curves crossing in $p$ is given by the sign of the cross product $(v_{a_j}\times v_{b_j})_z$, where the $3$-dimensional tangent vectors $v_{a_j}$ and $v_{b_j}$ lie in the direction of $a$ and $b$ in $p$, with their first component corresponding to the real direction and their second component to the imaginary direction (the third component being always zero).
The intersection number of $a$ and $b$ is defined as~\cite[Def. 14]{Bobenko2011}
\begin{equation}
a\circ b = \sum_{p\in a\cap b} (a\circ b)_p.
\end{equation}
The basis is referred to as canonical when 
\begin{equation}
a\circ a = 0,\ b\circ b = 0,\ a_j\circ b_k = \delta_{jk}.
\label{eq:canonicalbasis}
\end{equation}

For the torus, the homology basis consists of two cycles, one encircling the hole the short way, while the other goes around it the long way, see Fig.~\ref{fig:torus} (right).
The $A$- and $B$-periods are expressed in terms of Abelian integrals over the elements of the homology basis:
\begin{equation}
A_j = \int_{a_j} d\Omega,\quad B_j = \int_{b_j} d\Omega,\quad j=1,\ldots, g.
\label{eq:ABperiods}
\end{equation}
The Abelian integral $\Omega$ in~\eqref{eq:abelian_integral} takes an infinite number of different values at any point $p$ on the Riemann surface. However it takes a definite value at the point $p$ modulo the periods. The problem of inverting the Abelian integrals of the first kind is known as the Jacobi inversion problem. For $g=1$ the result is an elliptic function, that is, a doubly periodic function of one complex variable. 
As Jacobi recognized~\cite{Markushevich2006}, for $g>1$ the inversion must be carried out for all Abelian integrals of the first kind simultaneously. This leads to the Abelian or hyperelliptic functions, which are $2g$-periodic generalizations of elliptic functions and depend on $g$ complex variables.
The domain of the hyperelliptic functions is called the Jacobian variety and is given by the $g$-dimensional complex space quotiented by the period lattice generated by the periods of the holomorphic differentials~\eqref{eq:ABperiods}~\cite{Belokolos1994}. 
The mapping from a point on the Riemann surface into its Jacobian variety is known as the Abel map. 
The inversion of the Abel map provides the solution to the Jacobi inversion problem in terms of the Abelian functions. They are constructed using the Riemann theta functions. The inversion of the Abel map is sketched as part of the derivation of the exact solutions of the NLSE in Appendix~\ref{app:abelmap}.
The connection between nonlinear equations and theta functions is discussed in Ref.~\cite{Dubrovin1981}.
A useful guide to theory of Riemann surfaces in which most of the notions utilized in this Appendix are defined, is given by~\cite[Chapter 1]{Bobenko2011}.

\subsection{Governing equations}

The integrable, focusing NLSE~\cite{Turitsyn2017} in dimensionless form is given by
\begin{equation}
i\frac{\partial q(t,z)}{\partial z} + \frac{\partial^2 q(t,z)}{\partial t^2} + 2|q(t,z)|^2 q(t,z) = 0.
\label{eq:app:nlse}
\end{equation}

The integrability of this equation hinges on the fact that the NLSE can be obtained as the compatibility condition $\partial^2 \Phi(t,z,\lambda)/(\partial t \partial z) = \partial^2 \Phi(t,z,\lambda)/(\partial z \partial t)$ between two linear partial differential equations:
\begin{align}
\frac{\partial \Phi}{\partial t} &=  \left[-i\lambda\sigma_3 + \begin{pmatrix}0& q(t,z)\\-\bar{q}(t,z)&0\end{pmatrix}\right]\Phi=\!\!\mathop:U(t,z,\lambda) \Phi,\nonumber\\
\frac{\partial \Phi}{\partial z} &= \left[-2\lambda U + \begin{pmatrix} -iq(t,z)\bar{q}(t,z)&-\frac{\partial q(t,z)}{\partial t}\\-\frac{\partial \bar{q}(t,z)}{\partial t}&iq(t,z)\bar{q}(t,z),\end{pmatrix}\right]\Phi,\nonumber\\
\sigma_3 &= \begin{pmatrix}1&0\\0&-1\end{pmatrix}, \Phi(0,0,\lambda) = \begin{pmatrix}1&0\\0&1\end{pmatrix}.\label{eq:app:UV}
\end{align}
The condition holds true if and only if $q(t,z)$ is a solution to the NLSE.

The analytical expressions for finite-gap solutions are obtained as exact solutions to the partial differential equations \eqref{eq:app:derivative_of_q_app}~\cite[Eqs. 5.5, 5.6]{Kotlyarov2014} and \eqref{eq:app:derivative_of_mu_app}~\cite[Eqs. 5.7, 5.8]{Kotlyarov2014} below. These solutions can be expressed in an analytical form in terms of integrals over a Riemann surface. The derivation provided in this appendix is structured similarly to the derivation in~\cite{Tracy1988}. However, it is adapted to the optical channel, and provides in particular further details obtained from~\cite{Belokolos1994} and~\cite{Kotlyarov2014}.

The space-time dependence of any finite-gap solution is governed by the following differential equations~\cite{Kotlyarov2014}:
\begin{align}
\frac{\partial \log q(t,z)}{\partial t} =&2i\left(\sum_{j=1}^{g}\mu_j(t,z)-\frac{1}{2}\sum_{k=1}^{2g+2}\lambda_k\right),\notag\\
\frac{\partial \log q(t,z)}{\partial z} =&2i\left[\sum_{j>k}\lambda_j\lambda_k-\frac{3}{4}\left(\sum_{k=1}^{2g+2}\lambda_k\right)^2\right]\nonumber\\&-4i\left[\left(-\frac{1}{2}\sum_{k=1}^{2g+2}\lambda_k\right)\left(\sum_{j=1}^{g}\mu_j(t,z)\right)\right.\nonumber\\
&\left.\qquad+\frac{1}{2}\left(\left(\sum_{j=1}^g\mu_j(t,z)\right)^2-\sum_{j=1}^{g}\mu^2_j(t,z)\right)\right]\!\!.\label{eq:app:derivative_of_q_app}
\end{align}
The dynamics of the waveform $q(t,z)$ are essentially governed by the so-called auxiliary spectrum variables $\mu_j(t,z)$, which evolve on the Riemann surface $\Gamma$~\eqref{eq:app:riemann_surface}.

The motion of the auxiliary spectrum variables is governed by the differential equations~\cite{Kotlyarov2014}
\begin{align}
&\frac{\partial \mu_j(t,z)}{\partial t} = \frac{-2i\sigma_j\sqrt{\prod_{k=1}^{2g+2}(\lambda_k-\mu_j(t,z))}}{\prod_{l\neq j}(\mu_j(t,z)-\mu_l(t,z))},\nonumber\\
&\frac{\partial \mu_j(t,z)}{\partial z} = -2\left(\sum_{l\neq j}\mu_l(t,z)-\frac{1}{2}\sum_{k=1}^{2g+2}\lambda_k\right)\frac{\partial \mu_j(t,z)}{\partial t},
\label{eq:app:derivative_of_mu_app}
\end{align}
Their evolution is completely independent of the solution $q(t,z)$, but can be thought of as driving the evolution of $q(t,z)$ according to above equations.

The derivation essentially proceeds as follows. First, the differential equation for $\mu_j(t,z)$, Eq.~\eqref{eq:app:derivative_of_mu_app}, is integrated exactly. It is shown that the Abel-map, which was first introduced in algebraic geometry, linearizes the evolution of the auxiliary spectrum. 
Next, by using the Cauchy residue theorem, we obtain expressions for $\sum_j \mu_j(t,z)$ and $\sum_j \mu_j^2(t,z)$ in terms of integrals over the Riemann surface. Finally, the expressions for $\sum_j\mu_j(t,z)$ and $\sum_j \mu_j^2(t,z)$ are inserted into the differential equations~\eqref{eq:app:derivative_of_q_app} for $q(t,z)$ after which they can be solved analytically.
Finally, Sec.~\ref{app:baker} shows that the integration constant and two further parameters can be obtained as integrals over meromorphic differentials over the Riemann surface. 
If one is not interested in a specific auxiliary spectrum, it can be fixed implictly. The solution is thus obtained through integrals over a Riemann surface defined by the main spectrum.

In the derivation, several functions and integrals on the Riemann surface $\Gamma$ are utilized. 
For a point $p=(P,\lambda)$ on the Riemann surface, $\lambda(p)$ denotes the $\lambda$-coordinate of that point. Each value $\lambda$ corresponds to two different points on the surface $(+P,\lambda)$ and $(-P,\lambda)$. It is common to denote the points on the Riemann surface by their $\lambda$-coordinate, keeping the sign of $P$ implicit. If necessary we distinguish between the two points on the Riemann surface by superscript, $\lambda^\pm$.

\subsection{Integration of Eq.~\eqref{eq:app:derivative_of_mu_app}}

The Abel map $\underline{u}$ is defined for an arbitrary reference point $p_0 = (P_0,\lambda_0)$ on the Riemann surface as the following map from sets of points on $\Gamma$ to $\mathds{C}^g$:
\begin{equation}
\underline{u}_{p_0}\left(\{p_k\}\right) = \sum_{j=1}^n \left(\int_{p_0}^{p_j}d\psi_1,\ldots, \int_{p_0}^{p_j}d\psi_g\right)^T,\label{eq:app:abelmap}
\end{equation}
where $d\psi_j$ is a basis of normalized holomorphic differentials as defined in the paper. The notation $\{p_k\}$ is shorthand for $\{p_k|k\in 1,\ldots, n\}$, which denotes a set of $n$ arbitrarily numbered points on the Riemann surface. Here $n$ is arbitrary. Given a basis of holomorphic differentials defined on the Riemann surface $\Gamma$,
\begin{equation}
dU_j = \frac{\lambda^{j-1}}{P(\lambda)}d\lambda,\ j=1,\ldots,g,
\label{eq:app:diff}
\end{equation}
the normalized differentials are defined as
\begin{equation}
d\psi_j = \sum_{k=1}^g C_{j,k}\, dU_k,\label{eq:app:normdiff}
\end{equation}
where $\mathbf{C}=\mathbf{A}^{-1}$ and
\begin{equation}
A_{j,k} = \int_{a_k} dU_j.
\label{eq:app:matrixA}
\end{equation}

We denote by $\underline{W}(t,z)$ the Abel-map of the auxiliary spectrum:
\begin{equation}
\underline{W}(t,z) = \underline{u}_{p_0}(\{\mu_j(t,z)\}).
\end{equation}
$\underline{W}$ evolves linearly as a function of $t$ and $z$~\cite{Kotlyarov2014}. This is proven by computing the derivative of $W_j$ with respect to either $t$ or $z$, as below, and showing that it is constant.
Using Eq.~\eqref{eq:app:normdiff}, the components $W_j(t,z)$ of $\underline{W}$ are computed explicitly as:
\begin{align}
W_j(t,z) &= \sum_{k=1}^g \int_{p_0}^{\mu_k(t,z)}d\psi_j \nonumber\\ &=\sum_{k=1}^g\sum_{l=1}^g C_{j,l}\int_{p_0}^{\mu_k(t,z)}\frac{\lambda^{l-1}d\lambda}{P(\lambda)}.
\end{align}
The time-derivative of this vector is given by: 
\begin{equation}
\frac{\partial}{\partial t}W_j(t,z) =\sum_{k=1}^g\sum_{l=1}^gC_{j,l}\frac{\mu_k^{l-1}(t,z)\frac{\partial \mu_k(t,z)}{\partial t}}{P(\mu_k(t,z))}.\nonumber
\end{equation}
Inserting the derivative of $\mu_k(t,z)$ from Eq.~\eqref{eq:app:derivative_of_mu_app}, this is found to be equal to:
\begin{equation}
\frac{\partial}{\partial t}W_j(t,z) = \sum_{l=1}^g- 2iC_{j,l} \sum_{k=1}^g \frac{\mu_k^{l-1}(t,z)}{\prod_{n\neq k}(\mu_k(t,z)-\mu_n(t,z))}.\nonumber
\end{equation}
The Cauchy residue theorem~\cite{Bronstein2004} equates the integral of an analytic function over a closed path in the complex plane to the sum of the
 residues of its poles inside the closed path. Therefore the above sum over $k$ can be rewritten as a contour integral:
\begin{align}
I_0 &= \sum_{k=1}^g \frac{\mu_k^{l-1}(t,z)}{\prod_{n\neq k}(\mu_k(t,z)-\mu_n(t,z))}\nonumber\\
&=\frac{1}{2\pi i}\int_C \frac{\lambda^{l-1}d\lambda}{\prod_{n=1}^g(\lambda-\mu_n(t,z))},
\end{align}
where $C$ is a contour in the complex plane that encloses the poles $\mu_k$. 

On the other hand this integral can also be evaluated by reversing the direction of the contour, which picks up the residue at $\lambda=\infty$. The latter can be done by change
of variables $v = 1/\lambda$, which transforms the differential according to $dv =  -\lambda^{-2}d\lambda$. Multiplying the numerator and denominator by $v^g$ yields the following result:
\begin{equation}
I_0 = \frac{1}{2\pi i}\int_{C'} \frac{v^{g-l-1} }{\prod_n^g (1-\mu_n v)} dv.
\end{equation} 
The integrand only has a pole at $v=0$ for $l=g$. Evaluating the residue at $v=0$, the integral becomes equal to the Kronecker-delta $\delta_{l,g}$. Therefore the derivative of $W_j$ with respect to $t$ equals
\begin{equation}
\frac{\partial}{\partial t}W_j(t,z) = -2iC_{j,g} =\!\!\mathop: \frac{1}{2\pi}\omega_j.
\label{eq:app:dWdt}
\end{equation}
The matrix $\mathbf{C}$ is given by integrals over the Riemann surface according to~\eqref{eq:app:matrixA} and therefore depends on the main spectrum only. Hence it is constant and $W_j$ is a linear function of $t$.

A similar, but more involved computation~\cite{Tracy1984} allows one to evaluate the $z$-derivative of $\underline{W}$:
\begin{align}
\frac{\partial}{\partial z}W_j(t,z) &= -4i\left[C_{j,g-1}+\frac{1}{2}\left(\sum_{k=1}^{2N+2}\lambda_k\right)\, C_{j,g}\right] \nonumber\\&=\!\!\mathop: \frac{1}{2\pi}k_j.
\label{eq:app:dWdz}
\end{align}
Integrating Eq.~\eqref{eq:app:dWdt} and Eq.~\eqref{eq:app:dWdz} gives

\begin{equation}
\underline{W}(t,z) = \frac{1}{2\pi}(\underline{\omega} t +\underline{k}z+\underline{d}),
\label{eq:app:W}
\end{equation}
where the integration constant $\underline{d}$ is determined by the Abel-map of the initial value of the auxiliary variables: 
\begin{equation}
\frac{\underline{d}}{2\pi}	 = \underline{u}_{p_0}(\{\mu_j(0,0)\}).
\label{eq:app:ddef}
\end{equation}

\subsection{Abel map and inversion of the Abel map}
\label{app:abelmap}

For given $t$ and $z$ the auxiliary spectrum is obtained by inverting the Abel map. This is known as the Jacobi-inversion problem and can be stated as follows: For a point $\underline{V}$ in the Jacobian variety of $\Gamma$ find the set $\{p_j\}$ such that $\underline{u}_{p_0}(\{p_j\})= \underline{V}$. Jacobi formally solved this problem by constructing an analytic function whose zeroes are precisely the points $p_j$.
He showed that this function is defined by~\cite{Bobenko2011}
\begin{equation}
F(p) = \theta\left(\underline{u}_{p_0}(\{p\})-\underline{K}+\underline{V}|\bm{\tau}\right),
\label{eq:app:abelinvert}
\end{equation}
where
\begin{equation}
K_j = \frac{1}{2}\tau_{jj} - \sum_{k=1}^g\int_{a_k}d\psi_k(p')\int_{p_0}^{p'}d\psi_j(p)
\label{eq:app:Kdef}
\end{equation}
and the $g\times g$ period matrix $\bm{\tau}$ is defined as
\begin{equation}
\tau_{ij} = \sum_{k=1}^g (A^{-1})_{i,k} B_{k,j}\textrm{,\,\,\, } B_{k,j} = \int_{b_j} dU_k.
\label{eq:app:tau}
\end{equation}

Note that here the argument of the Abel map $\underline{u}$ only consists of a single point $p$. 
The vector $\underline{K}$ as well as the Abel map depend on the base point $p_0$. However the derivatives of $F$ with respect to $p$ as well as its roots are independent of $p_0$. This is also true for its Taylor expansion in the neighborhood of any of its roots. Since $F$ is analytic, it is equal to its Taylor expansion and hence independent of $p_0$.
In the last equation and in the the remainder of the Appendix we therefore suppress the dependence of the Abel map on $p_0$. 

Now define a function $F(t,z,p)$, by:
\begin{equation}
F(t,z,p) = \theta\left(\underline{u}(\{p\})-\underline{K}+\underline{W}(t,z)|\bm{\tau}\right),\label{eq:app:Ffunction}
\end{equation}
where the Abel map of the auxiliary spectrum~\eqref{eq:app:W} is inserted in place of $\underline{V}$ in the defining equation of $F$, Eq.~\eqref{eq:app:abelinvert}. The function $F$ formally solves the differential equation for $\mu_j(t,z)$: The $\mu_j(t,z)$ appear as zeroes of $F(t,z,p)$.

\subsection{Evaluating the sums over $\mu_j(t,z)$}

It is important that the differential equations for $q(t,z)$, Eq.~\eqref{eq:app:derivative_of_q_app}, only depend on the $\mu_j(t,z)$ as sums $\sum_j \mu_j(t,z)$ and $\sum_j \mu_j^2(t,z)$, since these sums can be rewritten as contour integrals over $\Gamma$ by means of the Cauchy residue theorem. The procedure is similar to the evaluation of $I_0$ above. We first construct the contour integrals and evaluate them explicitly in terms of loop integrals over the Riemann surface.

Since F has zeroes at $\mu_j(t, z)$, $\frac{1}{F}d\lambda = \frac{d\log(F)}{d\lambda}d\lambda \equiv d\log(F)$ has first order poles at $\mu_j(t, z)$ with residue 1. Therefore a meromorphic differential with the residues $\mu_j(t,z)$ is given by $\lambda(p) d\log F(t,z,p)$. Similarly, $\lambda^2(p) d\log F(t,z,p)$ has residues with value $\mu_j^2(t,z)$. The factors $\lambda$ and $\lambda^2$ have introduced extra poles at $p=\infty^\pm$, whose residue must be subtracted, as shown below.

In order to apply the residue theorem, a contour that encloses all the poles must be obtained. One way to construct such a contour is to cut the Riemann surface along the cycles in the homology-basis and to straighten out the resulting surface to obtain a polygon~\cite[Sec. 1.3.2]{Bobenko2011}. This so-called canonical dissection is detailed in Ref.~\cite{Tracy1984} and yields a simply connected surface $\Gamma^*$. This surface contains all points of the original Riemann surface exactly once and hence all the poles $\mu_j(t,z)$.
It can be shown that the boundary $\partial\Gamma^*$ of $\Gamma^*$ is given in terms of the homology basis as $a_1b_1a_1^{-1}b_1^{-1}\ldots a_gb_ga_g^{-1}b_g^{-1}$, where e.g. $a_1^{-1}$ denotes the same cycle as $a_1$, but traversed in opposite direction.
 
The integrals that need to be evaluated therefore are the following:
\begin{align}
I_1 = \frac{1}{2\pi i}\int_{\partial\Gamma^*}\lambda\, d\!\log F(t,z,p),\label{eq:app:I_1}\\
I_2 = \frac{1}{2\pi i}\int_{\partial\Gamma^*}\lambda^2\, d\!\log F(t,z,p).
\label{eq:app:I_2}
\end{align}
Based on the residue theorem the $\sum_j\mu_j(t,z)$ and $\sum_j\mu_j^2(t,z)$ are found to be:
\begin{align}
\sum_j \mu_j(t,z) =  I_1&- \text{Res}_{\infty^+} [\lambda(p)\, d\!\log F(t,z,p)] \nonumber\\&- \text{Res}_{\infty^-}[\lambda(p) \, d\!\log F(t,z,p)],\label{eq:app:muandI1}\\
\sum_j \mu^2_j(t,z)= I_2&- \text{Res}_{\infty^+} [\lambda^2(p)\, d\!\log F(t,z,p)] \nonumber\\&- \text{Res}_{\infty^-}[\lambda^2(p)\, d\!\log F(t,z,p)].\label{eq:app:muandI2}
\end{align}
We exemplify the evaluation of the residue $\lambda(p)\, d\!\log F(p)$ at $p = \infty^+$. The other computations proceed in the same way.

First we write $\lambda\, d\log F = \lambda\, (d\log F/d\lambda) d\lambda$ and apply the chain rule:
\begin{align}
&\lambda(p)\, d\!\log F(t,z,p) = \sum_{j=1}^g\left(\lambda(p)\frac{\partial \log F(t,z,p)}{\partial u_j} \frac{du_j({p})}{d\lambda}\right) d\lambda\nonumber\\
&=\sum_{j=1}^g\sum_{k=1}^g\left(\lambda(p)\frac{\partial \log F(t,z,p)}{\partial u_j}\frac{C_{j,k}\lambda^{k-1}(p)d\lambda}{P(\lambda)}\right), 
\end{align}
where have used that the argument of the function $F$, Eq.~\eqref{eq:app:Ffunction} only depends on $\lambda$ only through the Abel map, Eq.~\eqref{eq:app:abelmap} and that its differentiation with respect to $\lambda$ yields $C_{j,k}\lambda^{j-1}/P(\lambda)$ by virtue of the definition of the normalized differentials, Eqs.~\eqref{eq:app:diff} and~\eqref{eq:app:normdiff}.

 In the limit of $\lambda\to\infty$, $1/P(\lambda)$ becomes $\lambda^{-g-1}+\mathcal{O}(\lambda^{-g-2})$, which means the previous sum approaches:
	\begin{align}
&\lim_{p\to\infty^\pm}\lambda(p)\, d\!\log F(t,z,p)\nonumber\\
=&\lim_{p\to\infty^\pm}\sum_{j=1}^g\sum_{k=1}^g\left[D_j \log F(t,z,p)\right] C_{j,k}\lambda^{k-g-1}(p)	d\lambda.
\end{align}

The only term for which this expansion has a residue at $\lambda = \infty$ is $k=g$. In that case, the substitution $v = 1/\lambda$ is employed again, to find:
\begin{align}
\text{Res}_{\infty^+}[\lambda(p)\, d\!\log F(t,z,p)] = \sum_{j=1}^gC_{j,g}
\frac{\partial \log F(t,z,\infty^+)}{\partial u_j}.
\label{eq:app:residu}
\end{align}
The right hand side of~\eqref{eq:app:residu} is, up to a multiplicative factor, equal to the $t$-derivative of $F$:
\begin{align}
\frac{\partial}{\partial t}\log F(t,z,\infty^+) &= -\sum_{j=1}^g \frac{\partial \log F(t,z,\infty^+)}{\partial u_j}\frac{\partial W_{j}}{\partial t} \nonumber\\&= 2i \sum_{j=1}^g \frac{\partial \log F(t,z,\infty^+)}{\partial u_j}C_{j,g},
\label{eq:app:dxF}
\end{align}
where~\eqref{eq:app:dWdt} was used. Combining Eqs. \eqref{eq:app:residu} and \eqref{eq:app:dxF}, one obtains:
\begin{equation}
\text{Res}_{\infty^+}[\lambda(p)\, d\!\log F(t,z,p)] = -\frac{i}{2}\frac{\partial}{\partial t} \log F(t,z,\infty^+).
\end{equation}

Computing the other residues in a similar fashion and inserting them into Eqs.~\eqref{eq:app:muandI1} and~\eqref{eq:app:muandI2} one obtains:
\begin{align}
\sum_j \mu_j(t,z)  =& I_1- \frac{i}{2}\frac{\partial }{\partial t}\log\left[\frac{F(t,z,\infty^-)}{F(t,z,\infty^+)}\right],\label{eq:app:I11}\\
\sum_j \mu^2_j(t,z)=& I_2  +\frac{1}{4}\frac{\partial}{\partial t} \log [ F(t,z,\infty^+)F(t,z,\infty^-)]\nonumber\\&-\frac{i}{4}\frac{\partial }{\partial z}\log\left[\frac{F(t,z,\infty^-)}{F(t,z,\infty^+)}\right].\label{eq:app:I21}
\end{align}
In this equation $I_1$ and $I_2$ remain unknown. They are now evaluated by separating the border $\partial\Gamma^*$ into the constituent $a$- and $b$-cycles. If $F$ were single valued on the surface $\Gamma$ the contributions of $a$ and $a^{-1}$ would cancel.  However, $F$ is not single valued. While changing the argument $p$ by an $a$-cycle does not change the value of $F$,\footnote{Since the theta function has period 1, and the integral of the differentials $d\psi_j$ is precisely $1$ on the $a$-cycles.}changing the argument $p$ to $p'$ by moving it around a $b$-cycle changes the value of $F$ by
\begin{equation}
F(t,z,p') = \exp(-2\pi i (u_j(\{p\})-K_j)-\pi i\tau_{jj})F(t,z,p).
\end{equation}
Consequently, the value of $d\!\log F(p')$ is given by:
\begin{equation}
d\!\log F(t,z,p')= -2\pi i d\psi_j +  d\!\log F(t,z,p).
	\end{equation}
	In the path $\Gamma^*$, the cycles $a_j$ and $a_j^{-1}$ are separated precisely by the cycle $b_j$ for any $j$.
This leads to the following relations for the contributions of specific cycles:
\begin{align}
\int_{a_j}d\!\log F(t,z,p)+\int_{a^{-1}_j}d\!\log F(t,z,p') &=\int_{a_j} 2\pi i d\psi_j,\nonumber\\
\int_{b_j}d\!\log F(t,z,p)+\int_{b^{-1}_j}d\!\log F(t,z,p') &= 0.\nonumber
\end{align}
By applying these relations to $I_1$ and $I_2$, Eqs.~\eqref{eq:app:I_1} and~\eqref{eq:app:I_2}, we find the following equalities:
\begin{align}
\int_{\partial\Gamma^*} \lambda\, d\!\log F(t,z,p) = \sum_j \int_{a_j} 2\pi i \lambda d\psi_j,\nonumber\\
\int_{\partial\Gamma^*} \lambda^2\, d\!\log F(t,z,p) = \sum_j \int_{a_j} 2\pi i \lambda^2 d\psi_j.\nonumber
\end{align}
For $I_1$ and $I_2$ this yields:
\begin{align}
I_1 = \sum_{j=1}^g\int_{a_j} \lambda d\psi_j,\label{eq:app:I12}\\
I_2 = \sum_{j=1}^g\int_{a_j} \lambda^2 d\psi_j, \label{eq:app:I22}
\end{align}
which provide $I_1$ and $I_2$ in terms of integrals over the Riemann surface.
By inserting \eqref{eq:app:I12} and \eqref{eq:app:I22} into Eqs.~\eqref{eq:app:I11} and~\eqref{eq:app:I21}, this yields an exact expressions for $\sum_j\mu_j(t,z)$ and $\sum_j\mu_j^2(t,z)$.

\subsection{Integration of Eq.~\eqref{eq:app:derivative_of_q_app}}
\label{app:integrateq}

The expressions for $\sum_j \mu_j(t,z)$ and $\sum_j\mu_j^2(t,z)$ obtained in the previous section are now inserted into Eq.~\eqref{eq:app:derivative_of_q_app} to obtain:
\begin{align}
\frac{\partial \log q(t,z)}{\partial t} &= \frac{\partial \log J(t,z)}{\partial t}+i\omega_0,\label{eq:app:logqt}\\
\frac{\partial \log q(t,z)}{\partial z} &= \frac{\partial \log J(t,z)}{\partial z}+ik_0,\label{eq:app:logqz}
\end{align}
where $\omega_0$ is given by
\begin{equation}
\omega_0=2I_1-\sum_{k=1}^{2g+2}\lambda_k, 
\label{eq:app:omega0def}
\end{equation}
and we have defined $J(t,z) := F(t,z,\infty^+)/F(t,z,\infty^-)$.
The derivative of $\log J$ in Eq.~\eqref{eq:app:logqz} comes from the second line in \eqref{eq:app:I21}. The other terms in Eq.~\eqref{eq:app:I21} and Eq.~\eqref{eq:app:derivative_of_q_app} are gathered in $k_0$. Even though the constituents of $k_0$ depend on $t$ and $z$, $k_0$ has been proven to be constant~\cite{Kotlyarov2014}, see also Sec.~\ref{app:baker}. The explicit expression can be found in the original derivation in the appendix of~\cite[below Eq. A30]{Tracy1988}.

By integrating Eqs. \eqref{eq:app:logqt} and \eqref{eq:app:logqz} simultaneously and inserting the definition of $F$, Eq.~\eqref{eq:app:Ffunction}, the analytical form of finite-gap solutions to the NLSE $q(t,z)$ is obtained:
\begin{align}
q(t,z) &= K_0\frac{\theta(\underline{r}^- -\underline{K}+\underline{W}(t,z)|\bm{\tau})}{\theta(\underline{r}^+ -\underline{K}+\underline{W}(t,z)|\bm{\tau})} e^{i\omega_0 t+ik_0z}\nonumber\\
&=K_0\frac{\theta\left(\frac{1}{2\pi}(\underline{\omega}t+\underline{k}z+\underline{\delta}^-)|\bm{\tau}\right)}{\theta\left(\frac{1}{2\pi}(\underline{\omega}t+\underline{k}z+\underline{\delta}^+)|\bm{\tau}\right)} e^{i\omega_0 t+ik_0z}.\label{eq:app:thetasol_app}
\end{align} 
Here 
\begin{equation}
\underline{r}^\pm=\underline{u}(\{\infty^\pm\})
\label{eq:app:rpmdef}
\end{equation}
is the Abel map evaluated at $p = \infty^\pm$.
The second line of this relation is found by inserting \underline{W} from Eq.~\eqref{eq:app:W} and absorbing all constant terms in the phases $\underline{\delta}^\pm$ according to
\begin{equation}
\frac{1}{2\pi}\underline{\delta}^\pm = \underline{r}^\pm-\underline{K}+\underline{d},
\label{eq:app:deltapmdef}
\end{equation}
which, gathering Eqs.~\eqref{eq:app:ddef},~\eqref{eq:app:Kdef} and~\eqref{eq:app:rpmdef} yields
\begin{align}
\frac{1}{2\pi}\delta^\pm_j =& \int_{p_0}^{\infty^\pm}d\psi_j -\frac{1}{2}\tau_{jj} + \sum_{k=1}^g\int_{a_k}d\psi_k(p')\int_{p_0}^{p'}d\psi_j(p)
\nonumber\\&- \sum_{k=1}^g\int_{p_0}^{\mu_k(0,0)}d\psi_j\label{eq:app:deltapm}.
\end{align}

For given main spectrum and initial value $|q(0,0)|$ (the phase of $q(0,0)$ is arbitrary), the expression~\eqref{eq:app:thetasol_app} with $\omega_j$ and $k_j$ given by Eqs.~\eqref{eq:app:dWdt} and~\eqref{eq:app:dWdz}, respectively, $\omega_0$ given by~\eqref{eq:app:omega0def}, $k_0$ given in~\cite[below Eq. A30]{Tracy1988} and the phases $\underline{\delta}^{\pm}$ by~\eqref{eq:app:deltapmdef}, in principle completely describes the corresponding finite-gap solution. However, its evaluation is inconvenient for a number of reasons.
Firstly, given an initial value $q(0,0)$, a compatible initial auxiliary spectrum must be obtained from the constraint described in Appendix A of the paper, which involves a numerical search procedure. 
Secondly, because $K_0$ appears in~\eqref{eq:app:thetasol_app} as an integration constant, its value depends on the initial condition $q(0,0)$. Note that while $K_0$ appears to depend on the auxiliary spectrum through the above mentioned constraint and~\eqref{eq:app:deltapm}, it is in fact independent of the auxiliary spectrum. Note also that the phase of $K_0$ is arbitrary (see Sec. III in the paper). Furthermore, $k_0$ is independent of $t$ and $z$, while as mentioned above, the explicit expression for it contains several parts that explicitly depend on $t$ and $z$ and involves derivatives with respect to these variables.

It turns out that finite-gap solutions can be obtained in a way that is computationally much simpler, given one is not interested in a specific initial auxiliary spectrum. In this case, the solution is obtained by providing the main spectrum and by computing integrals over the associated Riemann surface.
It has been shown~\cite[Eq. 4.3.22]{Belokolos1994} that any valid initial condition for the auxiliary spectrum corresponds to a phase $\underline{\delta}^+$ with vanishing imaginary part. Therefore, the auxiliary spectrum can be fixed implicitly by choosing, e.g., $\underline{\delta}^+=0$. According to~\eqref{eq:app:deltapm}, $\underline{\delta}^-$ can then be obtained from an integral over the Riemann surface
\begin{equation}
\frac{1}{2\pi}(\delta^+_j-\delta_j^-) = \int_{\infty^-}^{\infty^+}d\psi_j .
\label{eq:app:deltadiff}
\end{equation}
It remains to determine $\omega_0$, $k_0$ and $|K_0|$. In the following section it is shown that these parameters can be computed in terms of integrals over meromorphic differentials over the Riemann surface $\Gamma$.

\subsection{Computation of $\omega_0$, $k_0$ and $|K_0|$}
\label{app:baker}

The derivation provided below  follows the derivation in Sections 4.1 to 4.3 of~\cite{Belokolos1994}, but the notation is adapted to the optical channel. 
We proceed as follows: First, the most general expansion of a solution to the scattering problem, $\Phi(t,z,\lambda)$ is introduced, and it is shown how to extract $q(t,z)$ from this expansion. Next, a method is provided to obtain exact solutions to the scattering problem, Eq~\eqref{eq:app:UV}. Finally, a method for computing $|K_0|$, $\omega_0$ and $k_0$ is derived.

Let $\Phi(t,z,\lambda)$ denote a matrix-valued solution to the coupled equations~\eqref{eq:app:UV}. From these it follows that the matrix $U$ is given by
\begin{equation}
U(t,z,\lambda) = \frac{\partial\Phi(t,z,\lambda)}{\partial t}\Phi^{-1}(t,z,\lambda).
\label{eq:app:comp_of_U}
\end{equation}
It is shown in Ref.~\cite{Belokolos1994} that any exact solution $\Phi(t,z,\lambda)$ can be expanded around $\lambda =\infty$  in the following way:
\begin{equation}
\Phi(t,z,\lambda) =  \left[I+\sum_{k=1}^\infty\Phi_k(t,z)\lambda^{-k}\right]e^{-i\lambda t\sigma_3-2i\lambda^2z\sigma_3}C(\lambda),
\label{eq:app:Psiseries}
\end{equation}
where $I$ is the identity matrix and $C(\lambda)$ is an arbitrary invertible matrix.

The highest order of $\lambda$ in the derivative of $\Phi$ is $\mathcal{O}(\lambda)$, so it is necessary to compute the inverse of $\Phi^{-1}$ up to first order in $\lambda^{-1}$ in a neighborhood around $\lambda=\infty$:
\begin{align}
&\Phi^{-1}(t,z,\lambda) \nonumber\\&= C^{-1}(\lambda)e^{i\lambda t\sigma_3+2i\lambda^2z\sigma_3}\left[I-\Phi_1(t,z)\lambda^{-1}\right] + \mathcal{O}(\lambda^{-2}).
\end{align}
We also compute the $t$-derivative of $\Phi$, up to the constant term:
\begin{align}
&\frac{\partial \Phi}{\partial t} \nonumber\\&=-i\lambda\left[I+\Phi_1(t,z)\lambda^{-1}\right]\sigma_3e^{-i\lambda t\sigma_3-2i\lambda^2z\sigma_3}C(\lambda) + \mathcal{O}(\lambda^{-1}).
\end{align}
By multiplying the above two results, an expansion of $U(t,z,\lambda)$ is obtained near $\lambda=\infty$:
\begin{align}
U(t,z,\lambda)&= \frac{\partial \Phi}{\partial t}\Phi^{-1} \nonumber= -i\lambda\sigma_3-i\Phi_1\sigma_3+i\sigma_3\Phi_1+O(\lambda^{-1}),\nonumber\\
 &= -i\lambda\sigma_3+i[\sigma_3,\Phi_1]+\mathcal{O}(\lambda^{-1}).\label{eq:app:Uasymptotic}
\end{align}
where the square brackets denote the commutator. In the limit $\lambda\to\infty$ the $\mathcal{O}(\lambda^{-1})$-term vanishes, and ~\eqref{eq:app:Uasymptotic} becomes exact.
The commutator isolates the off-diagonal elements of $\Phi_1$, which yields
\begin{equation}
\lim_{\lambda\to\infty} U(t,z,\lambda)= \lim_{\lambda\to\infty}\left[-i\lambda\sigma_3 + \begin{pmatrix}0&2(\Phi_1)_{1,2}\\-2(\Phi_1)_{2,1}& 0\end{pmatrix}\right].\nonumber
\end{equation}
Comparing this $U$ in~\eqref{eq:app:UV}, the following relations for $q(t,z)$ are obtained:
\begin{equation}
q(t,z) = 2\left(\Phi_1(t,z)\right)_{1,2} = 2\left(\bar{\Phi}_1(t,z)\right)_{2,1}.
\label{eq:app:qfromPsi}
\end{equation}

For the remainder of this appendix, $t$ and $z$ can be considered fixed parameters.  A function is called meromorphic when it is holomorphic except on a set of isolated poles. $\Phi(t,z,\lambda)$ is meromorphic in the entire complex plane except at $\lambda = \infty$, where it has an essential singularity (as approaching it from different directions yields different values).
Therefore it is completely defined in the entire complex plane, through its poles and its asymptotic behavior around $\lambda = \infty$. Note that in the expansion~\eqref{eq:app:Psiseries}, the poles can still be chosen by changing the expansion coefficients $\Phi_k$.

 One way to make the choice of poles of $\Phi(t,z,\lambda)$, which is guaranteed to yield a finite-gap solution was provided in~\cite{Belokolos1994}  as follows:

Fix a main spectrum, and consider the Riemann surface $\Gamma$ corresponding to that main spectrum. There exists a unique function $\phi(t,z,p)$ with the following properties~\cite{Krichever1977,Belokolos1994, Kotlyarov2017}:
\begin{itemize}
\item $\phi(t,z,p)$ is a single-valued, meromorphic function on $\Gamma \backslash \infty^\pm$. The poles of $\phi(t,z,p) $ are given by the $g$ points $p_j$, $j=1,\ldots,p_g$ on $\Gamma$.
\item In the limit of $p\to\infty^\pm$, $\phi(p)$ is given by
\begin{align}
\lim_{p\to\infty^-} \phi(t,z,p) = [\begin{pmatrix}1\\0\end{pmatrix} + \mathcal{O}(\lambda^{-1})]e^{-i\lambda t -2i\lambda^2 z},\label{eq:app:asymptotics1}\\
\lim_{p\to\infty^+} \phi(t,z,p) = \lambda[\begin{pmatrix}0\\1\end{pmatrix} + \mathcal{O}(\lambda^{-1})]e^{i\lambda t +2i\lambda^2 z}.\label{eq:app:asymptotics}
\end{align}
\end{itemize}
The positions of the poles $p_j$ are constrained for the NLSE and correspond to a specific auxiliary spectrum. An analytic expression for $\phi(t,z,p)$ can be computed exactly, see Section 2.7 of Ref.~\cite{Belokolos1994}. For the computation of $k_0$ and $\omega_0$, we only need its asymptotic behavior.
$\phi(t,z,p)$ is a special case of a Baker-Akhiezer function, which is a single-valued function on a Riemann surface with a finite number of exponential singularities and poles. The components $\phi_i$, $i=1,2$ of the vector valued function $\phi$ can be separated into three parts; a meromorphic function $f_i(t,z,p)$ with poles at the points $p_j$, an exponential function $e^{R_i(t,z,p)}$, that provides the exponential divergence towards $\infty^\pm$ according to Eq.~\eqref{eq:app:asymptotics}, and a normalization constant $F_i(t,z)$, independent of $p$:  
\begin{equation}
\phi_i(t,z,p) = F_i(t,z) f_i(t,z,p)e^{R_i(t,z,p)}.\label{eq:app:phiin3}
\end{equation} 
Taking the limit $p\to\infty^\pm$ in Eq.~\eqref{eq:app:phiin3} and comparing to Eq.~\eqref{eq:app:asymptotics} gives:
\begin{equation}
F_1(t,z) = \frac{1}{f_1(t,z,\infty^-)},\quad F_2(t,z) = \frac{1}{f_2(t,z,\infty^+)}.
\end{equation}
Inserting this in Eq.~\eqref{eq:app:phiin3} we find
\begin{align}
\phi_1(t,z,p) &= \frac{f_1(t,z,p)}{f_1(t,z,\infty^-)}e^{R_1(t,z,p)},\label{eq:app:phi11}\\
\phi_2(t,z,p) &= \frac{f_2(t,z,p)}{f_2(t,z,\infty^+)}e^{R_2(t,z,p)}.
\label{eq:app:phi1}
\end{align}
We now combine two copies of $\phi(t,z,p)$ as
\begin{equation}
\Phi(t,z,\lambda) = \begin{pmatrix}\phi_1 (t,z,\lambda^-) & \phi_1 (t,z,\lambda^+)\\ \phi_2(t,z,\lambda^-)&\phi_2(t,z,\lambda^+)\end{pmatrix}
\label{eq:app:Philambda}
\end{equation}

Considering the asymptotic behavior of this matrix according to~\eqref{eq:app:asymptotics} and comparing to~\eqref{eq:app:Psiseries} in this limit, we can read off the matrix $C(\lambda)$ as:
\begin{equation}
C(\lambda) = \begin{pmatrix}1 & 0\\0&\lambda\end{pmatrix}.
\end{equation}

Removing the essential singularity in~\eqref{eq:app:Psiseries} by multiplying with the inverse of the exponential and taking the limit $\lambda\to\infty$ we find that:
\begin{align}
&\lim_{\lambda\to\infty} \Phi_{1,2}(t,z,\lambda) e^{i\lambda t\sigma_3 + 2i\lambda^2 z\sigma_3},\notag\\
=&\lim_{\lambda\to\infty} \left[I + \Phi_1(t,z)\lambda^{-1}\right]C(\lambda),
\end{align}
since all other terms in the expansion vanish. 
To obtain $q(t,z)$ from~\eqref{eq:app:qfromPsi} we need $(\Phi_1(t,z))_{1,2}$, which we can read off from above equation as
\begin{align}
q(t,z) &= \lim_{\lambda\to\infty}2\Phi_{1,2}(t,z,\lambda)e^{-i\lambda t - 2i\lambda^2 z}\\ &= \lim_{p\to\infty^+}2\phi_1(t,z,p)e^{-i\lambda(p) t - 2i\lambda^2(p) z},
\end{align}
where in the second line we used~\eqref{eq:app:Philambda}.
Inserting the expansion for $\phi_1$ from Eq.~\eqref{eq:app:phi1}, and $q(t,z)$ from Eq.~\eqref{eq:app:thetasol_app} we find:
\begin{align}
\lim_{p\to\infty^+}2\frac{f_1(t,z,p)}{f_1(t,z,\infty^-)}e^{R_1(t,z,p)}e^{-i\lambda(p) t - 2i\lambda^2(p) z} \\=K_0\frac{\theta\left(\frac{1}{2\pi}(\underline{\omega}t+\underline{k}z+\underline{\delta}^-)|\bm{\tau}\right)}{\theta\left(\frac{1}{2\pi}(\underline{\omega}t+\underline{k}z+\underline{\delta}^+)|\bm{\tau}\right)} e^{i\omega_0 t+ik_0z} .
\end{align}
In order to determine $\omega_0$ and $k_0$, we equate the exponential terms in this equation:
\begin{equation}
e^{i\omega_0 t+ik_0z} =\lim_{p\to\infty^+} e^{R_1(t,z,p)}e^{-i\lambda(p) t - 2i\lambda^2(p) z}.
\label{eq:app:expcomparison}
\end{equation}
Formally this has to be justified by computing the explicit expression for Eq.~\eqref{eq:app:phiin3}. It can be found in~\cite[Eq. 4.1.16]{Belokolos1994}.

By comparing~\eqref{eq:app:asymptotics1} with Eq.~\eqref{eq:app:phi11}, we can read off $R_1$ at $\infty^-$, but we need it at $\infty^+$. $R_1$ is a mermorphic function with poles only at $\infty^\pm$. That means it is completely defined by providing its Laurent series in a neighborhood of $\infty^-$.We can write $R_1$ in terms of meromorphic functions for which the  Laurent series is known at $\infty^-$ \emph{and} $\infty^+$ to obtain the value of $R_1$ at $\infty^+$.

For this purpose the following meromorphic functions $\Omega_j$  are introduced, via the meromorphic differentials\footnote{The $\Omega_j$ are defined via the differentials $d\Omega_j$, because the differentials are single valued on $\Gamma$, while their integrals are not.} $d\Omega_j$:
\begin{equation}
\Omega_j(p) = \int_{p_0}^p d\Omega_j. \label{eq:app:OmegaFromdOmega}
\end{equation}
The choice of $p_0$ does not influence the asymptotic behavior as long as $p_0$ is not a pole of $d\Omega_j$. The asymptotic behavior around $\infty^\pm$ of these meromorphic functions is chosen to reproduce the asymptotic behavior of $R_1$ and $R_2$:
\begin{align}
\Omega_0(p) &= \pm(\log \lambda + \mathcal{O}(1)), p\to\infty^\pm,\label{eq:app:divdiffs0}\\
\Omega_1(p) &= \pm(\lambda+\mathcal{O}(1)), p\to\infty^\pm,\label{eq:app:divdiffs1}\\ 
\Omega_2(p) &= \pm(2\lambda^2+\mathcal{O}(1)), p\to\infty^\pm.\label{eq:app:divdiffs2}
\end{align}
The meromorphic differentials $d\Omega_i$ are uniquely determined given that they have singularities at $\lambda=\infty^\pm$, and by the constraint that their integral over the $a$-cycles is zero (i.e., they are normalized)\cite[Theorem 15]{Bobenko2011}.\footnote{A solution can also be derived without normalizing the $\Omega_j$ on the $a$-cycles. The resulting parametrization of the solution to the NLSE however is not the same as the one derived here.}

We observe that, in the limit $p\to\infty^-$, the following equality holds for $R_1$ (compare Eq.~\eqref{eq:app:asymptotics1} and~\eqref{eq:app:phiin3} and Eqs.~\eqref{eq:app:divdiffs0} to~\eqref{eq:app:divdiffs2}).
\begin{equation}
\lim_{p\to\infty^-} i\Omega_1(p)t + i\Omega_2(p)z = \lim_{p\to\infty^-} R_1(t,z,p)+\mathcal{O}(1).\nonumber
\end{equation}
Similarly for $R_2$ we find, after replacing $\lambda$ in~\eqref{eq:app:asymptotics} with $e^{\log \lambda}$:
\begin{equation}
\lim_{p\to\infty^+}\Omega_0(p)+i\Omega_1(p)t+i\Omega_2(p)z =  \lim_{p\to\infty^+} R_2(t,z,p) + \mathcal{O}(1).\nonumber
\end{equation}

Let the subleading terms of the differentials be given by\footnote{In principle, the constant contribution depends on the initial point $p_0$ in Eq.~\eqref{eq:app:OmegaFromdOmega} and is different for the two different limits $p\to\infty^\pm$. Since the final results only depend on the difference between the subleading terms at $\infty^\pm$, the definition provided here is sufficiently general, and implicitly fixes the base point $p_0$.}:
\begin{align}
\Omega_0(p) &= \pm(\log \lambda+\frac{A_0}{2} + \mathcal{O}(\lambda^{-1})), p\to\infty^\pm,\label{eq:app:defK0_app}\\
\Omega_1(p) &= \pm(\lambda + \frac{A_1}{2}+\mathcal{O}(\lambda^{-1})), p\to\infty^\pm,\label{eq:app:defomega0_app}\\ 
\Omega_2(p) &= \pm(2\lambda^2+\frac{A_2}{2} + \mathcal{O}(\lambda^{-1})), p\to\infty^\pm.\label{eq:app:defk0_app}
\end{align}

It remains to relate the constants $A_i$ to parameters of the theta function solution.
The $\Omega_i$ defined in Eqs.~\eqref{eq:app:divdiffs0} to~\eqref{eq:app:divdiffs2} contain a nonzero constant contribution, while the constant contribution in the $R_i$ is zero. 
In the limit of $p\to\infty^-$, $R_1$ is equal to:
\begin{align}
&R_1(t,z,p)\nonumber\\
&=i\Omega_1(p) t + i\Omega_2(p)z+\frac{iA_1}{2}t+\frac{iA_2}{2}z+\mathcal{O}(\lambda^{-1}).
\end{align}

Because $R_1(t,z,p)$ is meromorphic, the same expansion holds at $\infty^+$ giving
\begin{align}
&\lim_{p\to\infty^+} R_1(t,z,p) \nonumber\\
=&\lim_{p\to\infty^+}\left[i\lambda(p) t + 2i\lambda^2(p)z\right]+iA_1t+iA_2z.\label{eq:app:exp_of_phi}
\end{align}
Inserting this result into Eq.~\eqref{eq:app:expcomparison} yields
\begin{equation}
e^{i(\omega_0 t+k_0 z)} = e^{i(A_1t+A_2z)},
\end{equation}
which identifies $\omega_0 = A_1$ and $k_0 = A_2$.

$A_0$ is determined by repeating the same computation for $(\Phi_1)_{2,1}$, for which we obtain equations for both $q$ and $\bar{q}$:
\begin{align}
q(t,z)&= 2\lim_{p\to\infty^+}\frac{f_1(t,z,p)}{f_1(t,z,\infty^-)}e^{i(\omega_0 t+k_0 z)}, \nonumber\\
\bar{q}(t,z)&= 2\lim_{p\to\infty^-}\frac{f_2(t,z,p)}{f_2(t,z,\infty^+)}e^{-A_0-i(\omega_0 t-k_0z)}.\label{eq:app:qandqbar}
\end{align}

Note that the symmetry of the main spectrum under complex conjugation guarantees that $A_0$, $A_1$ and $A_2$ are real~\cite[p. 111]{Belokolos1994}. If $A_1$ or $A_2$ were not real, this would cause divergences in above equation.

The derivation of the exact form of the functions $f_1$ and $f_2$ is rather involved. It is given in~\cite[Sec. 2.7]{Belokolos1994}.
By inserting them into~\eqref{eq:app:qandqbar} and equating $q(t,z)$ and the conjugate of $\overline{q}(t,z)$ the following relation is obtained:
\begin{align}
&\frac{K_0}{2}\frac{\theta\left(\frac{1}{2\pi}(\underline{\omega}t+\underline{k}z+\underline{\delta}^+-\underline{r})|\bm{\tau}\right)}{\theta\left(\frac{1}{2\pi}(\underline{\omega}t+\underline{k}z+\underline{\delta}^+)|\bm{\tau}\right)} \nonumber\\&=-\frac{2}{\overline{K}_0}\frac{\overline{\theta}\left(\frac{1}{2\pi}(\underline{\omega}t+\underline{k}z+\underline{\delta}^++\underline{r})|\bm{\tau}\right)}{\overline{\theta}(\frac{1}{2\pi}(\underline{\omega}t+\underline{k}z+\underline{\delta}^+)|\bm{\tau})}e^{-A_0}.
\end{align}
Here $\underline{r} = \underline{\delta}^+-\underline{\delta}^-$.
By careful consideration of the symmetry of the main spectrum and symmetry of the period matrix $\bm{\tau}$, it is shown in section 4.3 of~\cite{Belokolos1994} that:
\begin{equation}
\overline{\theta}(\underline{v}|\bm{\tau}) = \theta(\underline{\overline{v}}|\bm{\tau}).
\end{equation}
Inserting this in the above equation, and utilizing that $\Im(\underline{\omega})=\Im(\underline{k})=0$ gives:
\begin{align}
\frac{K_0}{2}\frac{\theta\left(\frac{1}{2\pi}(\underline{\omega}t+\underline{k}z+\underline{\delta}^+-\underline{r})|\bm{\tau}\right)}{\theta(\frac{1}{2\pi}(\underline{\omega}t+\underline{k}z+\underline{\delta}^+)|\bm{\tau})} =\nonumber\\
-\frac{2e^{-A_0}}{\overline{K}_0}\frac{\theta(\frac{1}{2\pi}(\underline{\omega}t+\underline{k}z+\overline{\underline{\delta}}^++\overline{\underline{r}})|\bm{\tau})}{\theta(\frac{1}{2\pi}(\underline{\omega}t+\underline{k}z+\overline{\underline{\delta}}^+)|\bm{\tau})}.\label{eq:app:thetaconstraint}
\end{align}
This equality can only hold for all $t$ and $z$ when all prefactors are equal, and the theta function ratios are equal. For the amplitude $K_0$ this yields:
\begin{equation}
|K_0|^2 = -4e^{-A_0}.
\label{eq:app:amplitude}
\end{equation}

Finally, the integral over the differential is expressed in terms of the following limits:
\begin{equation}
\int_{\infty^-}^{\infty^+}d\Omega_k = \lim_{p\to\infty^+}\Omega_k(p)-\lim_{p\to\infty^-}\Omega_k(p).
\end{equation}
Inserting~\eqref{eq:app:defK0_app}-\eqref{eq:app:defk0_app} on the right-hand side of this equation, and using $A_0=-\log(-|K_0|^2/4)$, $A_1=\omega_0$, $A_2=k_0$, one obtains the following relations:
\begin{align}
-\log(-\frac{1}{4}|K_0|^2) &= \int^{\infty^+}_{\infty^-} d\Omega_0 - 2\int_1^\infty \frac{1}{\lambda} d\lambda,\label{eq:app:A0}\\
\omega_0 &= \int^{\infty^+}_{\infty^-} d\Omega_1-2\int_0^\infty d\lambda,\label{eq:app:A1}\\
k_0 &=\int^{\infty^+}_{\infty^-} d\Omega_2-2\int_0^\infty 4\lambda d\lambda.\label{eq:app:A2}
\end{align}
The last integral in each equation subtracts the leading divergence in the first. Its lower boundary is chosen such that it does not introduce a non-zero constant term.

For the vectors $\underline{\delta}$ and $\underline{r}$, Eq.~\eqref{eq:app:thetaconstraint} leads to the following constraints:
\begin{align}
\underline{\delta}^+ = \overline{\underline{\delta}}^+ + 2\pi\underline{N},\\
\underline{r} = -\overline{\underline{r}}+2\pi\underline{M}.
\end{align}
where $\underline{N}$ and $\underline{M}$ are both arbitrary vectors of integers. Due to the periodicity of the theta function, every vector $\underline{N}$ and $\underline{M}$ corresponds to the same solution of the NLSE. The second constraint can be rewritten as
\begin{equation}
\Re(\underline{r}) = \pi\underline{M}.
\end{equation} In practice this is already satisfied due to the symmetry of the main spectrum, see the analysis in \cite[Sec. 4.3.2]{Belokolos1994}.
It can be used to verify if $\underline{r}$ is properly computed. In the first constraint, an arbitrary value can be chosen for $\underline{N}$. For $N_j=0$, this gives the constraint $\Im(\underline{\delta}^+)=0$. From this derivation it can not be seen that all finite-gap solutions are obtained by choosing a main spectrum and a real vector $\underline{\delta}^+$. This can be seen from the original derivation in~\cite[Sec. 4.3.2]{Belokolos1994}.

The derivation in this Appendix provides the complete description of finite-gap solutions in terms of integrals over a Riemann surface defined by the main spectrum. 
To summarize, the analytical expression is given by Eq.~\eqref{eq:app:thetasol_app}, where the period matrix $\bm{\tau}$ is defined in~\eqref{eq:app:tau}, $\underline{\omega}$ in \eqref{eq:app:dWdt}, $\underline{k}$ in \eqref{eq:app:dWdz} and the phases are determined by \eqref{eq:app:deltadiff}, provided one sets $\underline{\delta}^+=0$. Finally, $|K_0|$, $\omega_0$ and $k_0$ are given by Eqs.~\eqref{eq:app:A0}-\eqref{eq:app:A2}.

\subsection{Constraint on the auxiliary spectrum}
\label{app:aux_spec}

Given a main spectrum $\lambda_k$, $k=1,\ldots,2g+2$, not every initial auxiliary spectrum $(\mu_j(0,0),\sigma_j(0,0))$, $j=1,\ldots,g$ and initial condition $q(0,0)$ correspond to a valid initial condition for the solution of the NLSE. The constraint to obtaining a solution to the NLSE was first given in~\cite{Tracy1984}, and is repeated here. 

Let $P(\lambda)$ be defined as in Eq.~\eqref{eq:defineP}. Define the function $f$ through:
\begin{equation}
f^2(\lambda)\mathop:\!\!=P^2(\lambda)-|q(0,0)|^2\prod_{j=1}^g (\lambda-\mu_j(0,0))(\lambda-\bar{\mu}_j(0,0)).
\end{equation}
Then the set of $\mu_j(0,0)$ correspond to a solution of the NLSE when $f(\lambda)$ is a polynomial of finite degree. The proof of this claim is available in Appendix 1 of~\cite{Tracy1984}. Note that this constraint does not provide an explicit expression to compute a valid auxiliary spectrum.

\subsection{Shift of spectrum}
One property which follows directly from the evolution equations for $q(t,z)$ and $\mu_j(t,z)$, Eqs.~\eqref{eq:derivative_of_mu} and~\eqref{eq:derivative_of_q}, is that a shift in the linear spectrum of a signal corresponds directly to a shift in the nonlinear spectrum.
This can be shown by taking a $\mu_j(0,0),\lambda_j$ and $q(0,0)$ which define a solution to the NLSE, and by considering the shifted spectrum: $\mu_j(0,0)\to\mu_j(0,0)+\Lambda$ and $\lambda_j\to\lambda_j+\Lambda$, with $\Lambda\in\mathds{R}$. The time derivative of $\mu_j(t,z)$~\cite[(5.8)]{Kotlyarov2014}, Eq.~\eqref{eq:derivative_of_mu}, is unchanged, showing that $\mu_j(t,0) \to \mu_j(t,0)+\Lambda$ for all $t$.  The time-derivative of $\log q$~\cite[(5.5)]{Kotlyarov2014}, Eq.~\eqref{eq:derivative_of_q} picks up an extra $-2i\Lambda$ term. Therefore, if $q(t,0)$ corresponded to the original signal, the new signal $q_\Lambda(t,0)$ is given by $q(t,0)e^{-2i\Lambda t}$.

\bibliographystyle{ieeetr}
\bibliography{GoossensPartI} 
\end{document}